\documentclass{ametsocV6.1}

\AddToHook{cmd/@maketitle/before}{%
\lhead{\parbox[b]{\textwidth}{\normalfont\small\color{gray}\centering
This work has been submitted for publication to Weather and Forecasting (WAF). Copyright in this work may be transferred without further notice.\par
\vspace{4pt}
Generated using the official AMS \LaTeX\ template v6.1\par}}
}

\renewcommand{\draftspace}{\renewcommand{\baselinestretch}{1}}
\nolinenumbers
\let\internallinenumbers\relax

\DeclareUnicodeCharacter{00D7}{\ensuremath{\times}}
\DeclareUnicodeCharacter{00B1}{\ensuremath{\pm}}
\DeclareUnicodeCharacter{00B0}{\ensuremath{^{\circ}}}
\DeclareUnicodeCharacter{2010}{-}
\DeclareUnicodeCharacter{2264}{\ensuremath{\leq}}

\title{Toward Actionable Predictability of Tropical Mesoscale Convective Systems through Large-scale Recurrent Environments in Northern South America.}

\authors{Vanessa Robledo\aff{a}\correspondingauthor{Vanessa Robledo, vanessa-robledodelgado@uiowa.edu},
John F. Mejía\aff{b},
Jhayron S. Pérez-Carrasquilla\aff{c},
Mohamed Abdelkader\aff{a},
Humberto Vergara\aff{a}}

\affiliation{\aff{a}{IIHR-Hydroscience \& Engineering, University of Iowa, Iowa City, IA}\\
\aff{b}{Division of Atmospheric Science, Desert Research Institute, Reno, NV}\\
\aff{c}{Department of Atmospheric and Oceanic Science, University of Maryland, College Park, MD}}

\abstract{
Mesoscale convective systems (MCSs) produce much of the extreme precipitation, floods, and landslides that affect northern South America (NOSA), yet remain difficult to anticipate because they are strongly governed by local processes. On the night of 31 March 2017, an MCS triggered a catastrophic flash flood in Mocoa, Colombia ---an event whose severity was not foreseen with sufficient lead time for emergency action, which typically requires several hours to a couple of days. Motivated by this disaster, this study investigates whether such events emerge within recurrent environments that provide a source of large-scale predictability, rather than as isolated and inherently unpredictable occurrences. We combine a 21-year (2001--2021) MCS tracking dataset, ERA5 reanalysis, equatorial Rossby (ER) wave diagnostics, and a Self-Organizing Map (SOM) to identify recurrent environments associated with MCS occurrence in the Orinoco--Amazon basin. The Mocoa environment maps onto one of nine recurrent configurations characterized by enhanced low-level moisture, neutral Orinoco low-level jet conditions, favorable upper-level circulation, and strong ER-wave activity. We further trace convective precursors to an upstream source region over the Guiana Highlands up to 24~h before arrival in the Mocoa region. Together, these results show that the Mocoa event was not climatologically exceptional but a high-impact expression of a recurrent, partly predictable large-scale regime, demonstrating that portion of MCS predictability arise from recurring large-scale conditions that can be monitored hours to days in advance, and support hazardous weather forecasting in NOSA and other tropical regions.}

\begin{document}

\maketitle


\statement
Mesoscale convective systems (MCSs) produce many of the floods, landslides, and extreme rainfall events that affect northern South America, yet forecasting them remains challenging because tropical convection is strongly influenced by local-scale processes. By examining the atmospheric conditions associated with the devastating 2017 Mocoa disaster and comparing them with two decades of MCS observations, we found that the event developed within a recurrent large-scale atmospheric pattern favorable for storm organization. More broadly, our analysis identifies a suite of favorable synoptic environments linked to enhanced moisture availability and large-scale atmospheric disturbances. Tracking these atmospheric patterns may provide actionable predictability on timescales of several hours to days, supporting earlier warnings of hazardous weather in tropical regions. 


\section{Introduction}

A large fraction of extreme precipitation in northern South America (NOSA) is associated with Mesoscale Convective Systems (MCSs) \citep{feng2021global, robledo2024climatological}. MCSs are complexes of thunderstorms that span at least 100 km and persist for several hours \citep{houze2018100}. These systems can be primary drivers of hydrometeorological hazards that, in combination with complex topography, limited adaptive capacity and socio-economic challenges, make the population in this region highly vulnerable to impacts from natural disasters. In NOSA, recurrent MCS-related hazards include floods, flash floods and landslides \citep{poveda2020high}, which threaten the livelihoods of millions of people and underscore the need for improved forecasting capabilities to support robust early warning systems and reduce regional vulnerability.

In tropical environments, MCS development is primarily governed by moist deep convection and local-scale processes, including topographic forcing, boundary-layer thermodynamics, and land-atmosphere interactions, which tend to limit its predictability \citep{Fritsch2001}. However, the development and organization of MCSs can also be strongly modulated by large-scale atmospheric precursors, with enhanced predictability emerging primarily under favorable configurations \citep{judt2020atmospheric}.
Key large-scale precursors to tropical deep convection include convectively coupled equatorial waves (CCEWs), such as Equatorial Rossby waves (ER), Kelvin waves, and mixed Rossby--gravity waves \citep{kiladis2009convectively, cheng2023mesoscale};  tropical easterly waves \citep{ocasio2026review}; northerly low-level surges during the boreal winter \citep{abdillah2021cold}, frontal systems \citep{amorim2015some, siqueiraetal2005}; and gravity waves occurring in the Monsoon regions or the passage of the Intertropical Convergence Zone (ITCZ) \citep{houze2004mesoscale, Marlton2019Meteorological}. These systems promote convective organization through enhanced moisture convergence, ascent, upper-level divergence, and, for some disturbances, increased low-level cyclonic vorticity \citep{houze2004mesoscale, wolding2020interactions,Tomassini2018Mesoscale, mapes2006mesoscale}. In addition, low level jets (LLJs) \citep{algarra2019} contribute to the transport of moisture and can enhance moisture convergence that regulate the organization and intensity of tropical MCSs \citep{Schumacher2020The}.

Because these large-scale systems evolve on mesoscale to synoptic timescales, they can modulate moist deep convection over periods ranging from several hours to days \citep{Schumacher2020The, judt2020atmospheric}, providing forecasts of opportunity, which are critical for warning systems. Although these precursors do not deterministically predict individual events, they condition the atmosphere toward states of enhanced convective probability. Identifying and characterizing these synoptic patterns therefore provides a pathway to extend predictability of MCS development beyond the limits imposed by local-scale processes in the tropics. 

The catastrophic March 2017 Mocoa event provides an  opportunity to examine how large-scale atmospheric patterns create favorable environments for MCS development, increasing the likelihood of high-impact events in NOSA. Located in the southern Colombian Andes, the city of Mocoa experienced an extreme flash flood and debris flow triggered by an intense MCS. The disaster devastated entire neighborhoods, resulting in significant loss of life and widespread destruction. According to the Colombian National Unit for Disaster Risk Management (UNGRD), the event left at least 335 fatalities, 400 injuries, 200 missing people and more than 20,000 affected. Although there have been other impactful flash floods in the region \citep{aristizabal2020definicion}, the Mocoa disaster was unprecedented in scale and impact, making it one of the most severe weather--related disasters in recent decades. 

Several studies have sought to understand the drivers of this event. For example, \citet{garcia2019dynamic} analyzed the kinematic behavior of the debris flow and the geomorphological characteristics of the involved basins with intense deforestation that conditioned the event. \citet{cheng2018characteristics} suggested that the preceding 2014--2016 El Niño event contributed through prolonged drought and reduced vegetation cover. More recently, \citet{martinez2024mesoscale} explored the atmospheric conditions preceding the event, highlighting the Orinoco Low level Jet (OLLJ;  \citet{builes2022}) as a key feature associated with the MCS initiation and development. However, it remains unclear how unique the Mocoa MCS was within the regional climatology and whether its large-scale environment exhibited characteristics with enhanced predictability. Addressing this gap provides an opportunity to improve our understanding of both the predictability of extreme MCS--driven events and the identification of forecasts of opportunity for high-impact weather in tropical regions.

We hypothesize that recurrent large-scale atmospheric configurations associated with moisture transport and dynamical forcing create preferred environments for MCS development in NOSA, and that the March 2017 Mocoa event occurred within one of these climatologically favored yet highly impactful states, thereby offering a window of enhanced predictability. To test this hypothesis, we use a 21--year satellite-based MCS tracking dataset generated with the ATRACKCS algorithm \citep{robledo2024climatological}, previously evaluated over NOSA and more recently included in a global intercomparison of MCS tracking algorithms \citep{feng2025mesoscale}. We combine this dataset with Self-Organizing Map (SOM) analysis. Particularly in the extratropics, SOMs have been widely used to identify recurrent synoptic-scale circulation patterns and weather regimes associated with MCSs and extreme precipitation \citep{song2019contrasting, song2021crucial, zhang2022revealing, du2024unveiling}. However, their application to tropical MCSs environments remain limited. With the Mocoa event serving as a motivating case, we investigate the synoptic-scale atmospheric patterns modulating MCS development and associated extreme precipitation in the Orinoco--Amazon basin. Specifically, we identify recurrent large-scale configurations linked to MCS occurrence, characterize the properties of MCSs within these environments, and assess whether the Mocoa event represents an anomalous case or a manifestation of a climatologically favored state.

The remainder of this paper is organized as follows: Section 2 describes the MCS dataset, reanalysis data, and methodological framework, including the SOM approach. Section 3 presents the main results, including the characterization of the Mocoa event, the identification of recurrent synoptic patterns, and the relationship between these patterns and MCS properties. Section 4 summarizes the findings and discusses their implications and limitations.

\section{Data and methods}\label{sec2}

\subsection{MCS dataset}
We use the 2001--2021 MCS dataset from \citet{robledo2024climatological}, generated with the ATRACKCS algorithm to track deep convection over NOSA (90$^\circ$ -- 60$^\circ$W, 20$^\circ$N -- 10$^\circ$S). The dataset is based on hourly 0.1$^\circ$ NCEP/CPC Level 3 Merged Infrared Brightness Temperature (MergeIR V1) observations \citep{janowiak2017ncep} and Integrated Multi‐satellitE Retrievals for NASA Global Precipitation Measurement (GPM) (IMERG V06B) precipitation estimates \citep{huffman2015nasa}. For the 2017 Mocoa event analysis (Section 3\ref{sec3a}), we use the more recent IMERG V07 product.

To track MCSs, Cold Cloud Systems (CCSs) are identified using a maximum brightness temperature threshold (Tb $\leq$ 225K) and a minimum area threshold (area $\geq$ 2,000 km$^2$), with selected CCSs having at least five pixels with minimum precipitation rate (Prate $\geq$ 2 mm h$^{-1}$). Then, an area-overlap method links CCSs with $\geq50\%$ overlap between consecutive hourly observations into continuous tracks; these are added to the database as MCSs when criteria are sustained for at least six hours (more details in \citet{robledo2024climatological}). ATRACKCS has been used in global inter-comparisons against other tracking algorithms \citep{feng2025mesoscale}, proving that it is able to correctly represent important features of convection such as the diurnal and seasonal cycle, spatial distribution and MCSs characteristics.

\subsection{Reanalysis data}
The hourly 0.25° × 0.25° fifth generation European Center for Medium‐Range Weather Forecasts (ECMWF) reanalysis \citep[ERA5; ][]{hersbach2020era5} is used to analyze the synoptic and mesoscale environments preceding MCS formation and evolution. While ERA5 covers the full MCSs dataset record, the analysis is focused on February--May to cover a seasonal window around the event of interest.

Fields used in this study include geopotential height, winds, specific humidity, divergence, vertical velocity, temperature, and convective available potential energy (CAPE). We also used 16 vertical levels from 925 hPa to 200 hPa. Similar to other studies, we chose these variables to cover the broad range of dynamic and thermodynamic processes involved in MCSs development and evolution \citep{https://doi.org/10.1002/asl.1152, RelationshipsbetweenLargePrecipitatingSystemsandAtmosphericFactorsataGridScale, muetzelfeldt_mcsconditions}.

ERA5 reanalysis is used due to its global coverage, fine temporal resolution, and dynamically consistent atmospheric fields derived from the assimilation of diverse observations. Compared with earlier reanalyses such as ERA-Interim, ERA5 exhibits reduced biases \citep{hersbach2020era5} and improved representation of key variables in convective environments and tropospheric moisture relative to independent observations \citep{Taszarek:2021,Johnston:2021}. It has been widely used to characterize the large-scale environments associated with MCSs \citep[e.g.,][]{lee2026characterization, song2022observed} and provides a robust framework for environmental analyses \citep{soci2024era5}. Although all reanalyses contain systematic biases associated with model physics and data assimilation, limitations remain at local scales and shortly before convective initiation, where rapid atmospheric changes may not be fully captured and convective instability may be overestimated \citep{WU2024107108}. These limitations are unlikely to substantially affect our analysis, which focuses on the larger-scale environments associated with MCSs development, rather than resolve small-scale variability within individual storms.

\subsection{MCSs atmospheric environment characterization}

\subsubsection{Standardized Anomalies}
To characterize the atmospheric environment leading the occurrence of MCSs, the hourly standardized anomalies were computed relative to the climatology. This approach is widely used in weather analysis and forecasting because it reduces background variability, allowing for a clearer depiction of atmospheric structures, their interactions, and the early identification of dynamically relevant signals \citep{junker2009assessing, qian2021review}. 

For each variable, the long-term mean and standard deviation were calculated at each grid cell, pressure level, and hour of the day using the 2001--2021 record restricted to February--May. The standardized anomaly (z-score) was then obtained by subtracting the corresponding hourly climatological mean from the raw value and dividing by the hourly climatological standard deviation. This normalization removes the strong diurnal cycle inherent to tropical convection and produces anomalies that are directly comparable across space, time, and variables with different units, facilitating a consistent assessment of their relative importance.

\subsubsection{Self-Organizing Maps (SOM) Analysis}
Self-Organizing Map (SOM) is an unsupervised neural network clustering technique that projects high-dimensional data onto a low-dimensional space while preserving the topological relationships among similar patterns \citep{kohonen2013essentials,kohonen1998self}. In atmospheric science, SOMs have been widely used to identify recurrent large-scale circulation patterns and weather regimes associated with extreme precipitation \citep{zhang2022revealing, swales2016examining}, MCSs \citep{song2019contrasting,song2021crucial,du2024unveiling}, and other high-impact weather events \citep{horton2015contribution, loikith2017characterizing}. Here, we employed the SOM analysis to identify recurrent large-scale environmental patterns during the February--May period in NOSA, and to evaluate the uniqueness of the Mocoa MCS environment.

A baseline climatology was first constructed using hourly standardized anomaly fields of meridional wind (925 and 500 hPa), specific humidity (925 hPa), and geopotential height (500 hPa), averaged from $-12$h up to each time step during February--May (2001--2021). Variables were selected to capture moisture transport and synoptic forcing while minimizing redundancy among highly correlated environmental fields. The temporal smoothing reduces high-frequency variability and emphasizes the underlying synoptic-scale patterns. To reduce the dimensionality of the dataset, we applied Principal Component Analysis (PCA), retaining the first 25 components, which explain approximately 70\% of the total variance. Additional components contributed little additional variance and primarily introduced noise (see Figure \ref{fA1}).

Using the Python miniSOM library \citep{vettigliminisom}, we trained the SOM with these principal components. Hyperparameters were selected through sensitivity analysis of multiple grid configurations (2×2, 2×3, 2×4, 3×3, and 4×4), hyperparameters, and random initializations using quantization error (QE), defined as the average Euclidean distance between each input and its winning node, and the topographic error (TE), which is a measure of map topology preservation \citep{wang2022linking}. The 3×3 configuration provided the best balance of low QE and TE, while remaining stable across random initializations and capturing distinct yet non-redundant patterns (Appendix A).

To assess whether a specific SOM node is MCS--favorable or MCS--inhibiting, two datasets were constructed. The first, referred to as the \textit{MCS dataset}, includes MCSs initiating times in the Orinoco--Amazon basin during February--May (2001--2021). This resulted in a total of 17,067 samples. To avoid duplication, we retained only one event per hour when multiple MCS initiations occurred simultaneously, ensuring that each time step contributes a single sample. The resulting \textit{MCS dataset} contains 14,258 samples. The second dataset, referred to as the \textit{NoMCS dataset}, was derived from the climatology by selecting time steps with no MCS initiation occurring within a ±12 h window. This constraint was applied to isolate non-convective conditions. The resulting dataset contains 14,247 samples.

For each sample in both the \textit{MCS dataset} and the \textit{NoMCS dataset}, the environmental variables were averaged over the period from $t_0-12h$ to $t_0$ (where $t_0$ is the time of MCS initiation for the \textit{MCS dataset} or the time label for the \textit{NoMCS dataset}). Both datasets were projected onto the previously derived PCA space to ensure consistency in dimensionality, and subsequently mapped onto the trained SOM to identify their corresponding best-matching nodes.

Finally, similar to \citep{cheng2023mesoscale}, we defined an Enrichment Factor (R) for each node $i$ as:

\[
R_i = \frac{P(node_i | MCS)}{P(node_i | NoMCS)}
\]

where $P(node_i|MCS)$ and $P(node_i|NoMCS)$ represent the probability of assigning a sample to $node_i$ given MCS and non-MCS conditions, respectively. Values of $R_i >1$ indicate environments more frequently associated with MCS events (MCS--favorable), while $R_i <1$ indicates environments more frequently associated with non-MCS conditions (MCS--inhibiting).

The statistical significance of the MCS preference at the node-level was assessed using a Monte Carlo permutation test (n = 10,000). At each iteration, MCS labels were randomly reassigned across the pooled dataset, while preserving both the total number of MCS events and the total sample count within each SOM node. This procedure generates a null distribution of $R_i$ under the assumption that MCS occurrence is independent of SOM structure. The p-value was computed as the proportion of simulated R values above or below the observed $R_i$ value. Nodes were classified as MCS--favorable when $R>1$ and $p<0.01$ and as MCS--inhibiting when $R<1$ and $p<0.01$.

\section{Results}
\subsection{The Mocoa MCS Event}\label{sec3a}
\subsubsection{Event overview}

\begin{figure}[t]
 \noindent\centerline{\includegraphics[width=25pc,angle=0]{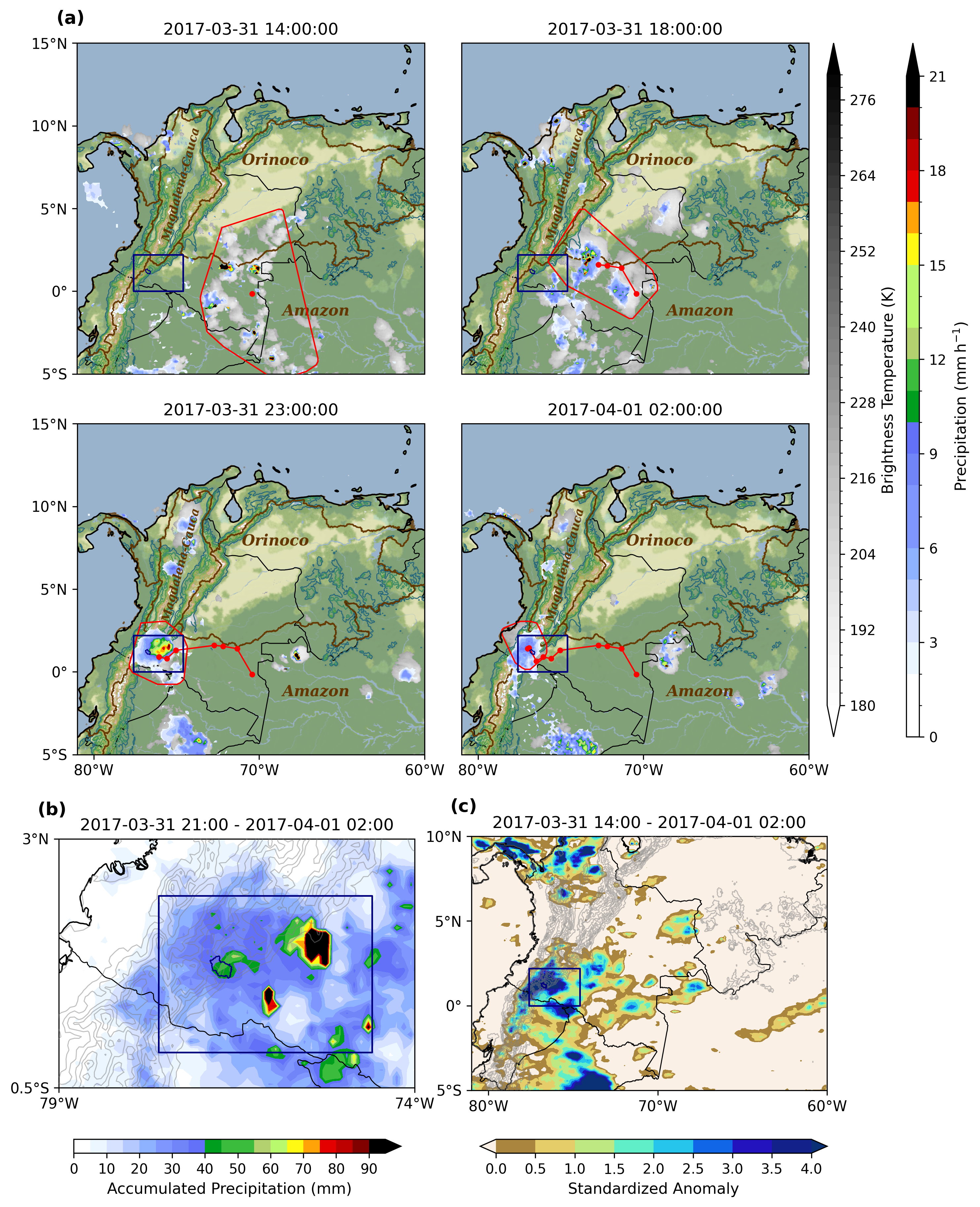}}\\
 \caption{Spatiotemporal analysis of the Mocoa MCS. (a) Mocoa MCS track at selected times (LT). Cloud-top brightness temperatures (BT $<$ 225 K) are shown in grayscale, with precipitation ($>$2 mm h$^{-1}$) overlaid in color. The red line denotes the MCS track produced by ATRACKCS, polygons delineate the MCS extent at each time step, and points indicate the MCS geometric centroid. The blue box indicates the Mocoa study region (MR), brown contours outline the main hydrologic basins, and elevation is shown for values above 550 m a.s.l. (b) Five-hour accumulated IMERG precipitation from March 31 at 2100 LT to April 1 at 0200 LT, 2017. The blue polygon delineates the affected Mocoa basin. (c) Standardized anomalies of accumulated precipitation from 1400 to 0200 LT, computed relative to the March--April climatology (2001--2021).}
\label{f1}
\end{figure}

\textbf{Figure \ref{f1}a} illustrates the trajectory of the Mocoa MCS identified by ATRACKCS. The system organized during the afternoon of 31 March 2017, over the Colombian Orinoco--Amazon region as a chain of connected cold convective clouds, some with high precipitation rates ($>15$ mm h$^{-1}$ according to IMERG V07). The MCS propagated westward, undergoing processes of merging, splitting, intensification and weakening before reaching the southern Colombian Andes at 2300 LT on March 31  where it intensified (lower-left panel in \textbf{Figure \ref{f1}a}).

According to IMERG, the maximum 5-h accumulated precipitation ($>100$~mm; March 31, 2100--April 1, 0200~LT) occurred northeast of Mocoa, along the eastern flank of the Eastern Cordillera of the Colombian Andes (\textbf{Figure \ref{f1}b}). Within the affected area (blue polygon in \textbf{Figure \ref{f1}b}), accumulated precipitation reached up to 55~mm, with a maximum precipitation rate of 19 mm h$^{-1}$ at 2300~LT on March 31. Available gauges near the Mocoa basin recorded maximum rates of 55~mm~h$^{-1}$ and a total event accumulation of 103.3~mm (not shown), ranking as the fourth-largest daily accumulation during 1984--2022 \citep{martinez2024mesoscale}.

To quantify how unusual the event was under the IMERG V07 lens, standardized anomalies of accumulated precipitation over 12-hour periods from March 31, 1400 LT to April 1, 0200 LT were computed relative to a March--April climatology (\textbf{Figure \ref{f1}c}). Precipitation anomalies during the MCS event reached $4\sigma$ above normal over the Mocoa region (hereafter MR; blue box). ATRACKCS tracked the system for 13 hours, estimating a mean velocity of 65 kmh$^{-1}$ and a convective area of approximately $120,000$ km$^{2}$ when the system was over Mocoa (2300 LT 31$^{st}$March). Additional ATRACKCS-derived characteristics are provided in Table \ref{t1}.

\subsubsection{Pre-convective environment at initiation}

\begin{figure}[b]
  \noindent\centerline{\includegraphics[width=39pc,angle=0]{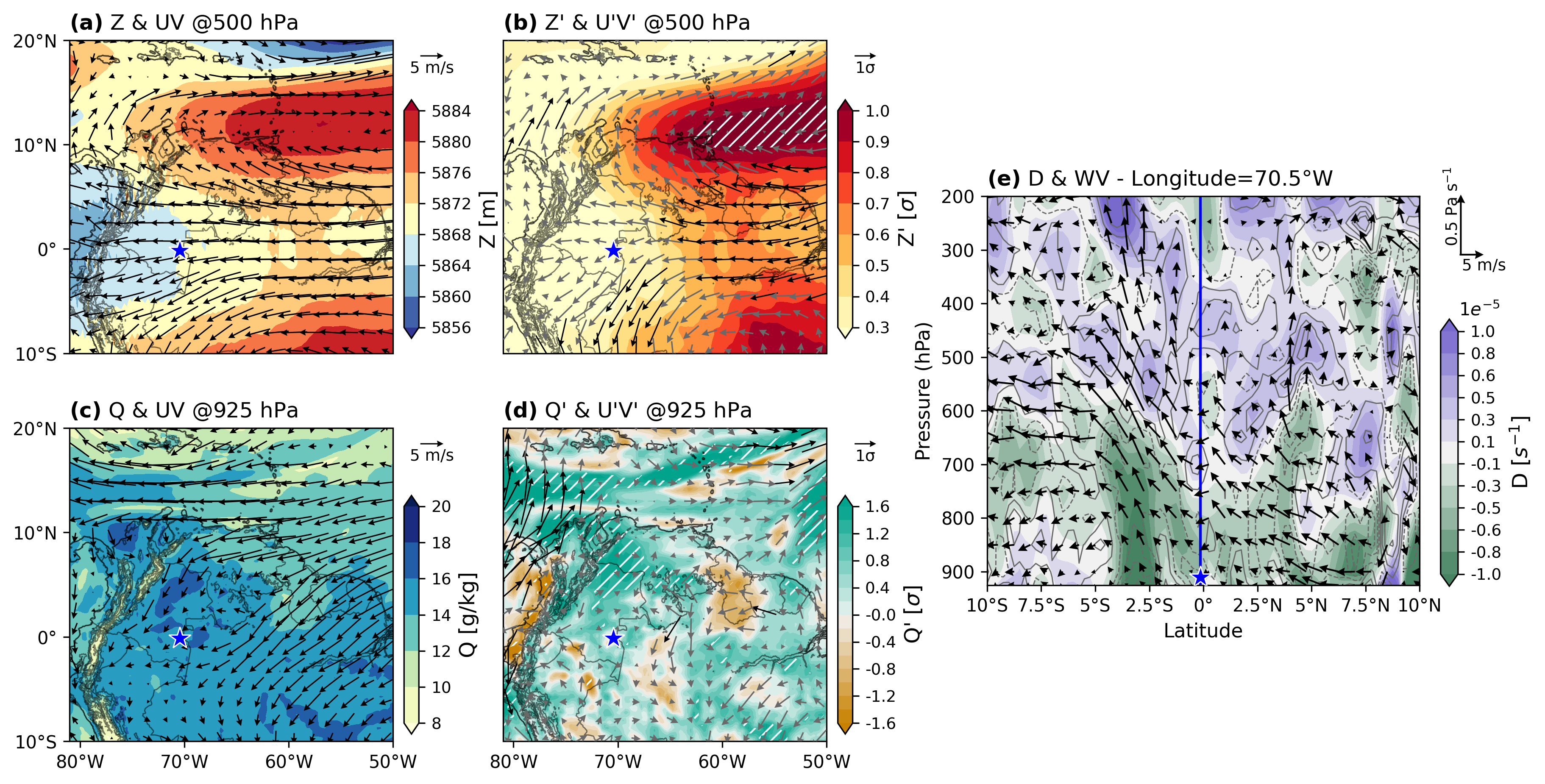}}\\
 \caption{Atmospheric precursors to the Mocoa event. Each panel displays 24-hour (March 30, 1400 LT -- March 31, 1400 LT) composites of: (a) Geopotential height (shading) and horizontal winds at 500 hPa, (b) Standardized anomalies (shading) of geopotential height and horizontal winds at 500 hPa, (c) Specific humidity (shading) and horizontal winds at 925 hPa, (d) Standardized anomalies of specific humidity and horizontal winds at 925 hPa, and (e) Vertical cross-section at 70.5$^{\circ}$W, divergence (shading) and standardized anomalies of divergence (dotted contours for negative ${\sigma}$ , continuous contour for positive ${\sigma}$, vectors denote meridional wind and vertical velocity. Black arrows and white hatching in (b) and (d) indicate regions where anomalies are statistically significant at the 90\% confidence level. Blue star indicates the geometric centroid of the MCS’s first detection (Figure 1a).}\label{f2}
\end{figure}

To carry out the analysis of the environment prior to the initiation of the Mocoa MCS, we consider -24 hours to -1 hour relative to initiation (t0 = 31 March, 1400 LT) to avoid the feedback from the system on the surrounding environment \citep{muetzelfeldt_mcsconditions}. \textbf{Figure \ref{f2}} shows composites of geopotential height (Z), specific humidity (q), and horizontal winds at 925 and 500 hPa, for both the raw fields and their corresponding hourly standardized anomalies. Here, the 10$^{th}$ and 90$^{th}$ percentiles of the hourly anomaly distribution (relative to the same hour as the Mocoa event) were used as thresholds to identify statistically rare conditions. Although not all anomalies exceed conventional significance thresholds, coherent spatial anomaly patterns are more likely to reflect organized atmospheric variability than isolated local departures.

At 925 hPa (\textbf{Figure \ref{f2}c--d}), specific humidity reached up to 18 g kg$^{-1}$ over the Orinoco--Amazon basin, with enhanced moisture near the initiation centroid of the Mocoa MCS (blue star) and standardized anomalies exceeding $1\sigma$. The horizontal wind field showed Atlantic inflow, forming what is known as the Orinoco Low-Level Jet (OLLJ) and the Guianas Jet. The OLLJ flows westward from 61$^\circ$W to 67$^\circ$W along 8--10$^\circ$N approximately, before being deflected equatorward and continuing along the eastern flank of the Cordillera from 8$^\circ$N to the equator \citep{builes2022}. The Guianas Jet is modulated by the topographic influence of the Guiana Highlands, flowing westward along their southern flank \citep{algarra2019}. Neither wind magnitude nor direction of the jets was anomalously strong 24 hours before the initiation of Mocoa MCS (\textbf{Figure \ref{f2}d}), and both jets appeared to be weaker-than-normal.

At 500 hPa, geopotential height and horizontal winds showed a westward-propagating pattern characterized by two geopotential height maxima north and south of the equator, with anticyclonic circulation around them (\textbf{Figure \ref{f2}a}). Standardized geopotential height anomalies exceeded $1\sigma$ over both maxima (\textbf{Figure \ref{f2}b}), while significant easterly wind anomalies (black arrows) over the northern Amazon followed the anticyclonic circulation. 

With these ingredients, the pre-initiation environment of the Mocoa MCS was characterized by a synoptic configuration resembling an equatorial Rossby (ER) wave, a convectively coupled equatorial wave characterized by westward-propagating disturbances associated with the organization of tropical convection \citep{kiladis2009convectively}. ER waves have previously been linked to change the likelihood of MCS occurrence in the tropics \citep{cheng2023mesoscale,nakamura2022aconvective,nakamura2022bconvective}, particularly over tropical South America, they can enhance MCS frequency by 20\%--50\% relative to climatology and generally elevate the probability of MCSs with extremely high rainfall \citep{cheng2023mesoscale}.

The two anticyclonic circulations (hereafter referred to as an ER-like pattern) induced mid- to upper-level divergence and low-level convergence in the region between the two gyres, near the equator, thereby favoring upward motion (\textbf{Figure \ref{f2}e}) and promoting convective initiation in a warm, moist, and unstable environment. This is supported by positive anomalies in CAPE ($>0.8\sigma$) and equivalent potential temperature ($\theta_e > 1\sigma$) (Figures \ref{fA3} and \ref{fA4}).

\begin{figure}[t]
\noindent\centerline{\includegraphics[width=19pc,angle=0]{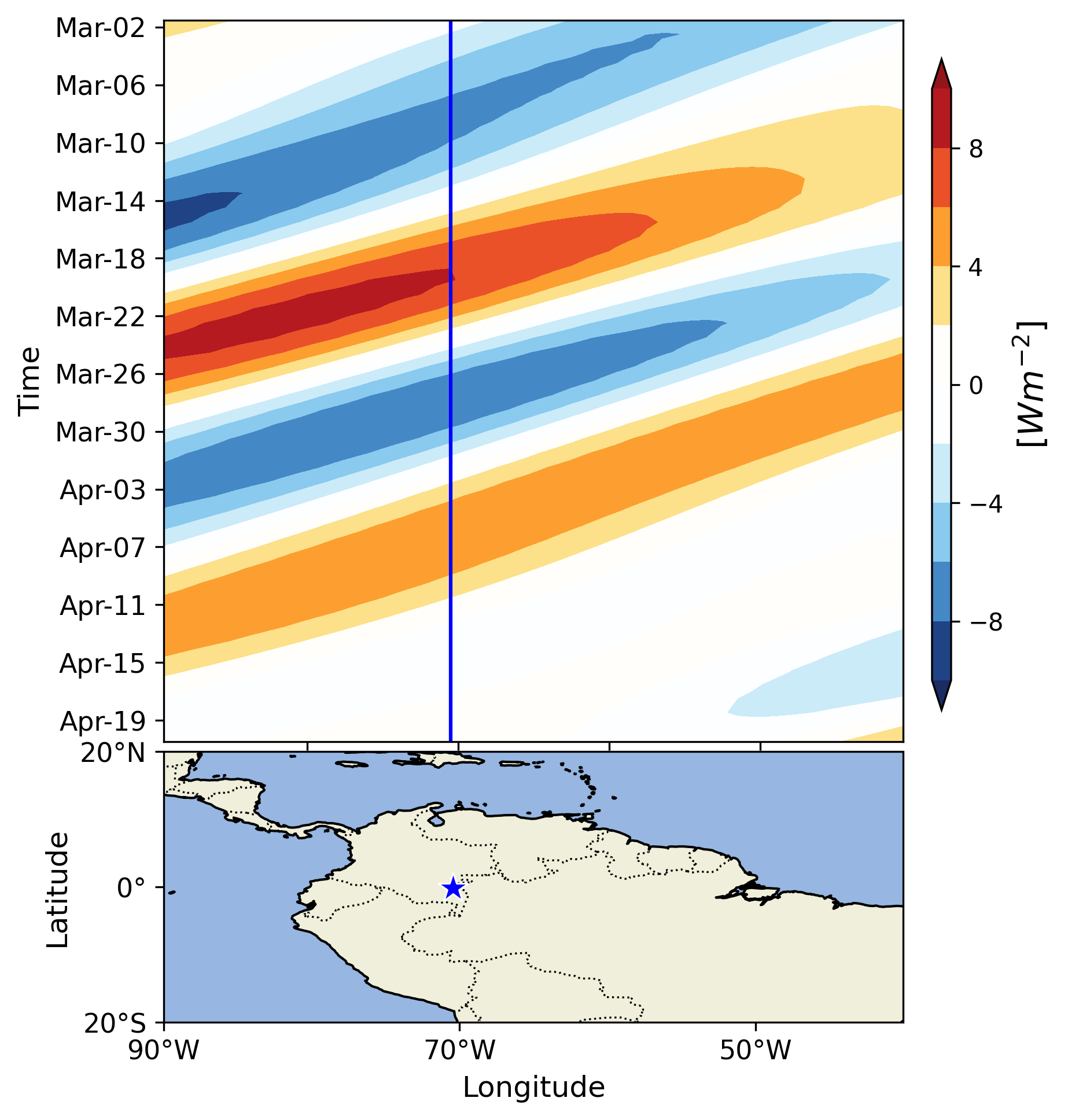}}\\
\caption{Hovmöller (longitude-time) diagram of ER-filtered OLR anomalies averaged between 20$^{\circ}$S and 20$^{\circ}$N. The filtering follows the \citet{wheeler1999convectively} methodology using \citet {matsuno1966quasi} wave dispersion characteristics.}\label{f3}
\end{figure}

To confirm the presence of an ER wave, we analyzed daily outgoing long wave radiation (OLR) data from National Oceanographic and Atmospheric Administration (NOAA) \citep{lee2025olr}, following the methodology from \citet{wheeler1999convectively} to isolate convectively coupled equatorial wave signals through wavenumber-frequency filtering, and applying the same filtering parameters for Matsuno modes \citep{matsuno1966quasi}. Our results (\textbf{Figure \ref{f3}}), confirmed the presence of a westward-propagating ER wave over tropical South America from March 29 to April 12, with its wet phase coinciding with the days preceding and following the initiation of the Mocoa MCS.

These findings are further supported by the operational diagnostics provided by the North Carolina Institute for Climate Studies\footnote{Available online at \url{https://ncics.org/portfolio/monitor/mjo/}}(see Figure \ref{fA5}). The same ER signal is identified in both OLR and 850-hPa zonal wind fields, indicating that the disturbance was vertically coupled to the mid-troposphere.

\subsubsection{Pre-intensification environment}\label{3_a_3}

\begin{figure}[t]
\noindent\centerline{\includegraphics[width=29pc,angle=0]{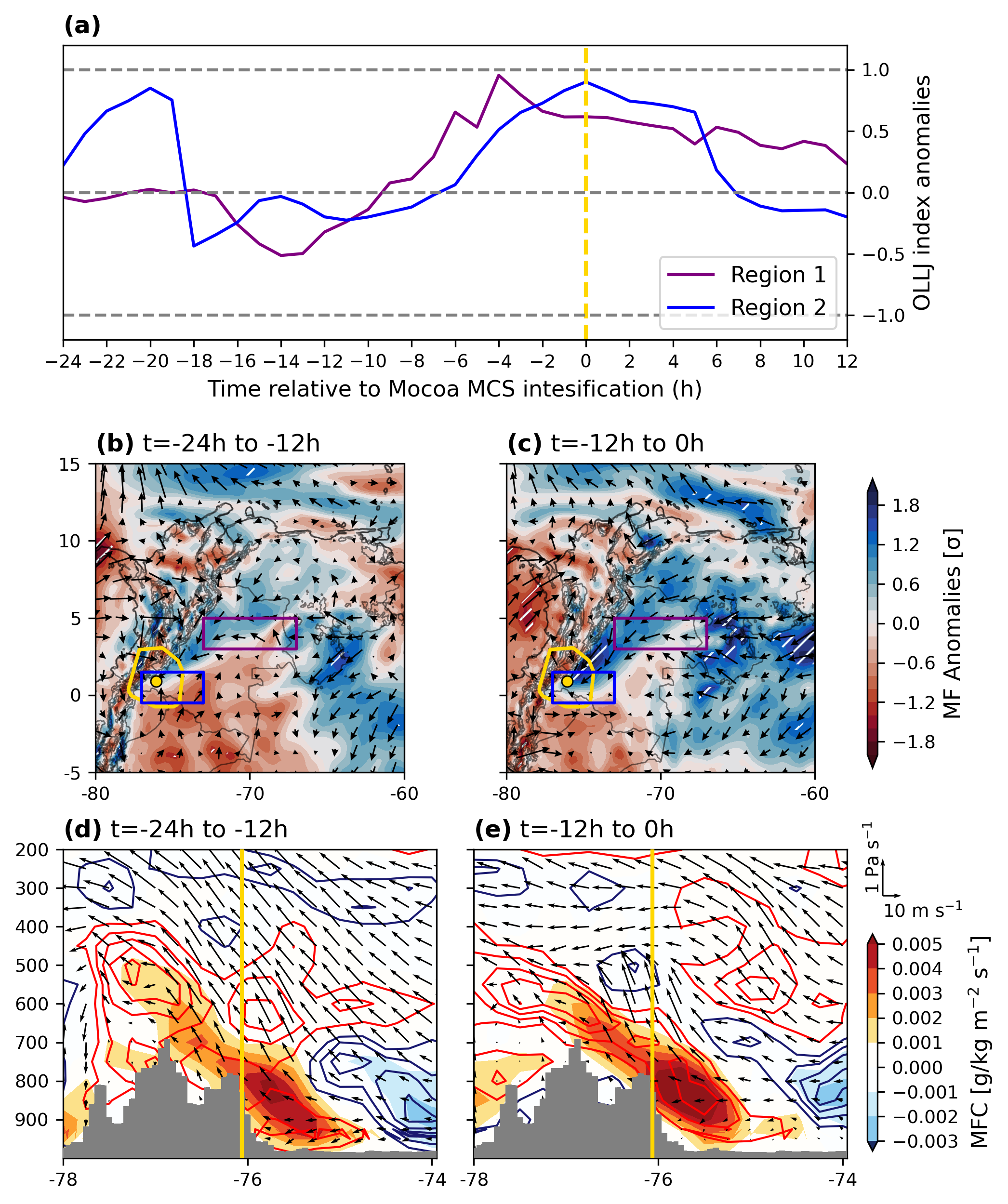}}\\
\caption{a) OLLJ index anomalies -24 hours prior the intensification of Mocoa MCS (t0, yellow dashed line) for region 1 (purple box) and region 2 (blue box), (b) and (c) Moisture flux anomalies -24 to 0h prior the intensification of Mocoa MCS, white hatching indicates regions where anomalies are statistically significant at the 90\% confidence level, yellow polygon and dot shows the event extend and its geometric centroid, (d) and (e) Vertical cross-section at 0.9${^\circ}$N latitude (same as Mocoa MCS centroid), moisture flux convergence (shading), red/blue contours positive/negative anomalies of moisture flux convergence every 0.5 ${\sigma}$, vertical velocity and zonal winds (black vectors) -24 to 0h prior the intensification of Mocoa MCS, yellow line representing Mocoa MCS centroid in longitude.}\label{f4}
\end{figure}

Once initiated, the MCS propagated toward the southern Colombian Andes, where it interacted with the complex topography. This interaction likely enhanced orographic lifting, leading to more vigorous ascent and higher precipitation rates compared to the earlier stages of the system.
However, this propagation pattern is not unusual for MCSs originating in the Orinoco--Amazon region where most systems travel westward reaching the eastern flank of the Andes \citep{robledo2024climatological}. The key question, therefore, is what factors contributed to the extreme intensity of this event upon reaching the topography. \citet{martinez2024mesoscale} attributed its intensity to an anomalously strong OLLJ.

To quantify the OLLJ evolution, we computed the hourly OLLJ index following \citet{correa2024drier}, defined as the spatial average of horizontal wind between 825 and 950~hPa over two regions. Region~1 (3$^\circ$--5$^\circ$N, 73$^\circ$--67$^\circ$W), over the Colombian Orinoco (purple box in \textbf{Figure \ref{f4}b,c}), and Region~2 (0.5$^\circ$S, 1.5$^\circ$N, 77$^\circ$--73$^\circ$W; blue box in \textbf{Figure \ref{f4}b, c}) corresponds to the jet exit region, at the transition between the Orinoco and Amazon basins and in proximity to Mocoa.

OLLJ index standardized anomalies were computed relative to February--May (2001--2021), with values above $1\sigma$ and below $-1\sigma$ indicating strong and weak jet conditions, respectively \citep{correa2024drier}. \textbf{Figure \ref{f4}a} shows the evolution of OLLJ index anomalies during the 24 hours preceding MCS intensification ($t_0$). The OLLJ began strengthening approximately 18 h before MCS intensification at $t_0$, although anomalies remained negative in both regions until around $-10$ h. In Region 1 (purple line), positive anomalies emerged around $-10$ h and reached a maximum intensity of $0.9\sigma$ at $-4$ h. In contrast, Region 2 (blue line) intensified later, with positive anomalies emerging near $-6$ h and peaking around $t_0$. Positive anomalies persisted in both regions after $t_0$ but gradually decreased over the subsequent 12 h.

As discussed by \citet{martinez2024mesoscale}, an active OLLJ was present during the night of the event. However, our results indicate that its magnitude did not exceed $1\sigma$ and therefore does not meet the conditions of a strong jet. Climatologically, the OLLJ exhibits a marked seasonal cycle, peaking during boreal winter (DJF), weakening in both intensity and spatial extent by March, and reaching its minimum strength during boreal summer (JJA) \citep{jimenez2019orinoco}. In addition, the OLLJ is a nocturnal jet characterized by a pronounced diurnal cycle, with peak intensity typically occurring between 0100 LT and 0700 LT \citep{builes2022}.

The standardized anomalies presented here explicitly account for these strong seasonal and diurnal cycles, allowing for a robust assessment of whether the jet conditions were anomalous relative to the expected background state. Our results show that the jet remained within neutral conditions, yet it still transported a substantial amount of moisture into the region (\textbf{Figure \ref{f4}b, c}) where moisture flux anomalies were up to $0.6\sigma$ from -14h to -12h and above $1.2\sigma$ from -12h to t0, coinciding spatially and temporally with the arrival of the MCS (yellow polygon in \textbf{Figure \ref{f4}b, c}) to the Andes.

The interaction between the moist flow and the complex topography likely enhanced orographic lifting and moisture flux convergence. \textbf{Figure \ref{f4}d, e} presents a vertical cross section at 0.9$^\circ$N (corresponding to the latitude of the MCS centroid), showing positive anomalies along the eastern flank of the Cordillera and over the MR. Persistent upward motion is evident throughout the 24 hours preceding the intensification, with the strongest ascent occurring between t=$-12$h and $t_0$. This alignment of dynamic and thermodynamic conditions may have contributed to the intensification of the system and the occurrence of high precipitation rates over the terrain.

\subsection{Climatological context of MCSs in the Orinoco--Amazon basin}

\subsubsection{Characteristics of MCSs}

\begin{figure}[t]
\noindent\centerline{\includegraphics[width=39pc,angle=0]{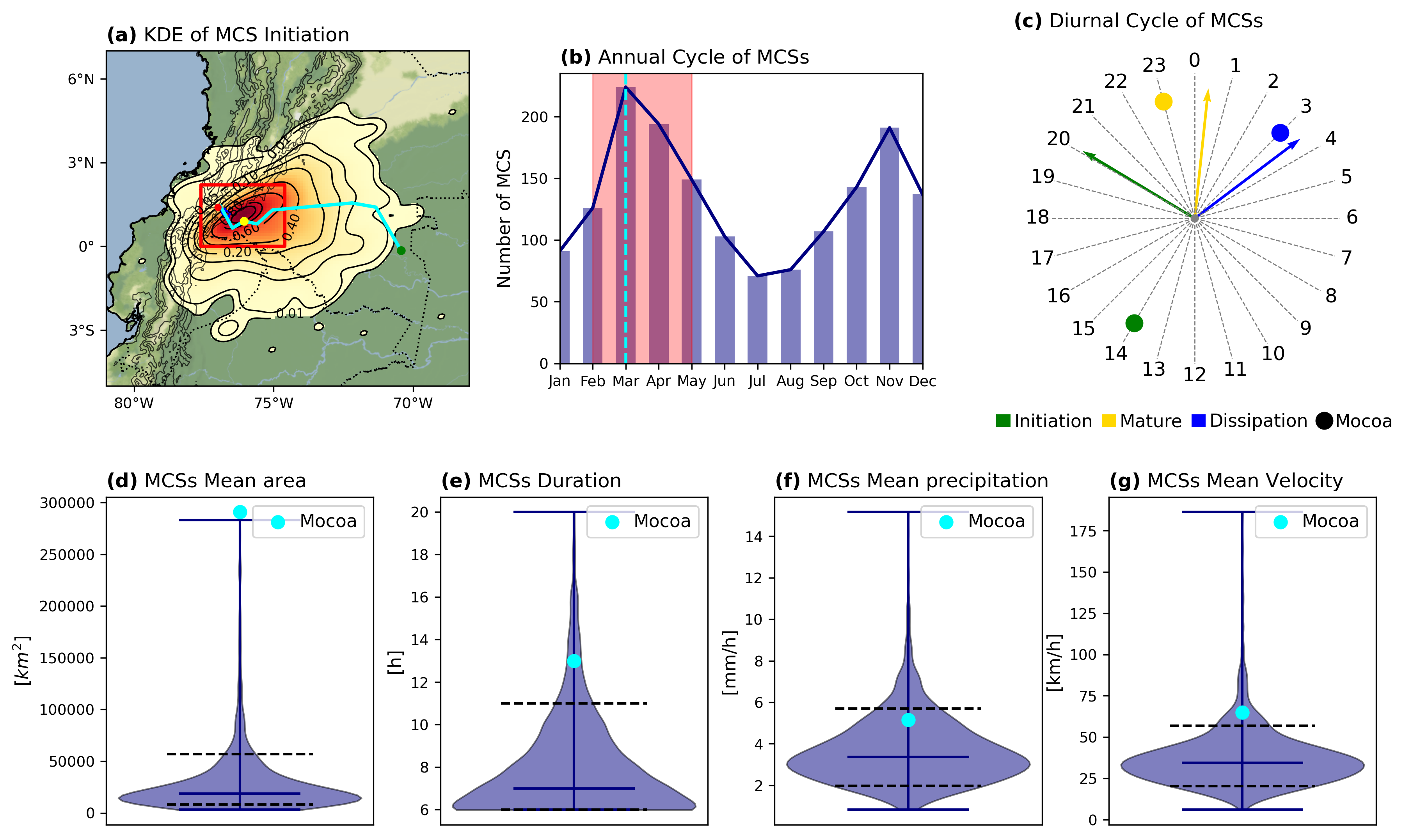}}\\
\caption{Climatological characteristics of MCS in Amazon--Orinoco basins. (a) Kernel density estimate (KDE) of MCS initiation locations. The red box is the Mocoa region (MR), Mocoa MCS track in cyan, and green, yellow, and red dots indicate its initiation, mature, and dissipation stages, respectively. (b) Annual cycle of the mean MCS frequency; the cyan line indicates the date of the Mocoa MCS. (c) First harmonic of the MCSs diurnal cycle.  Green, yellow, and blue arrows indicate the phase of maximum amplitude for the initiation, mature, and dissipation stages, respectively; dots indicate the corresponding phases of the Mocoa MCS. (d--g) Distributions of mean area, total duration, mean precipitation intensity, and mean propagation speed for MCSs passing through the Mocoa region (red box in panel a). Dashed black lines indicate the $10^{th}$ and $90^{th}$ percentiles, and cyan dots indicate values for the Mocoa MCS.}\label{f5}
\end{figure}

Characterizing MCSs in Mocoa region (red box in Figure \ref{f5}) provides a baseline to evaluate the predictability of these systems and to assess how anomalous was the Mocoa event within a historical context. \textbf{Figure \ref{f5}} illustrates the climatological characteristics of MCSs that passed over the study area, including their spatial initiation preferences (a), seasonal frequency (b), and diurnal lifecycle (c). Furthermore, panels (d) through (g) summarize some features of these systems, specifically mean area, duration, precipitation intensity, and propagation velocity, highlighting the position of the Mocoa event relative to the regional distribution.

The Mocoa event stands out for its spatial extent (\textbf{Figure \ref{f5}d}), with a mean area approaching 300,000 $km^2$, primarily attributed to the large area observed during the early stage of the event (see \textbf{Figure \ref{f1}}). In terms of temporal and dynamical characteristics, the event lies near the upper tail of the distribution, approaching the $90^{th}$ percentile for both duration and propagation velocity (\textbf{Figure \ref{f5}e, g}). Conversely, while its mean precipitation intensity exceeds the climatological median, it remains below the $90^{th}$ percentile according to IMERG. 

The annual cycle of MCSs in this region has a bimodal behavior (\textbf{Figure \ref{f5}b}), exhibiting peaks in March and November that align with the biannual passage of the ITCZ \citep{albrecht2016lightning}. The diurnal cycle is predominantly nocturnal, with MCSs typically initiating around 2000 LT, maturing near midnight, and dissipating around dawn. Although the Mocoa event followed the seasonal and diurnal pattern (dotted cyan line in \textbf{Figure \ref{f5}b}), its initiation timing preceded the climatological peak (\textbf{Figure \ref{f5}c}). This discrepancy is better understood by examining the regions of preferred initiation shown in \textbf{Figure \ref{f5}a}.

Initiation hotspots are concentrated along the eastern slopes of the Andes, where the spatial density of MCS initiation reaches its maximum (\textbf{Figure \ref{f5}a}). While this corridor represents the primary initiation zone, a secondary lower-density distribution of events extends across the Colombian Amazon and Orinoco basins, where the Mocoa MCS initiated. During its westward propagation, the Mocoa MCS underwent multiple cycles of merging, splitting, and intensity fluctuations. Such evolutionary changes, common among systems originating in the Orinoco--Amazon basin \citep{robledo2024climatological, greco1990rainfall}, challenge automated tracking algorithms based on fixed brightness-temperature or precipitation thresholds. Although the convective signal may weaken during these transitions, the underlying dynamic driver remains coherent. Consequently, tracking algorithms often break a single, long-lived system into multiple, independent tracks. This happens because the algorithm cannot distinguish between true system dissipation and temporary periods of reduced convective intensity, leading it to treat the system's evolution as the start of "new" events.

\begin{figure}[t]
\noindent\centerline{\includegraphics[width=39pc,angle=0]{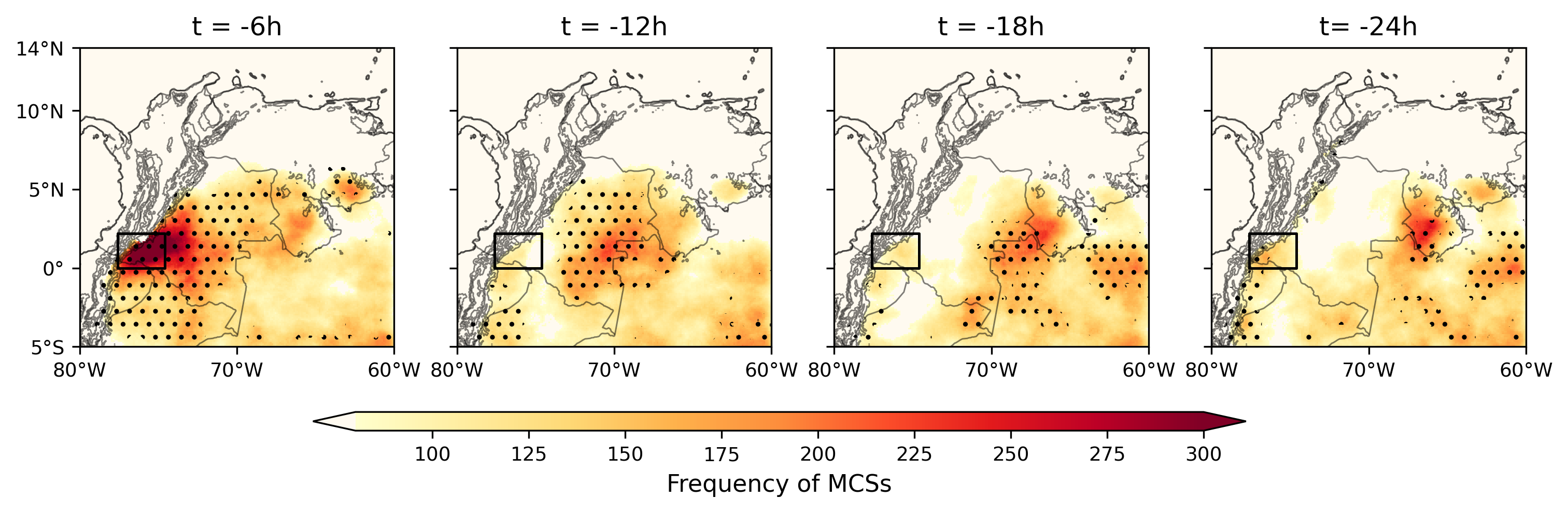}}\\
\caption{Lagged frequency of MCS occurrence 6, 12, 18, and 24 hours prior to MCS occurrence in Mocoa region (black box). Stippling (hatching) indicates regions where frequencies are statistically significant at the 99\% confidence level (p$<$0.01), based on a Monte Carlo test with 1,000 random iterations.}\label{f6}
\end{figure}

To address this limitation, we performed a lagged frequency analysis to identify the location of convective activity before MCSs reached the MR. Specifically, we examined the spatial distribution of MCSs 6, 12, 18, and 24 hours before an MCS is detected in MR. \textbf{Figure \ref{f6}} shows a clear temporal evolution consistent with westward MCS propagation. At $-$6 hours, MCSs were primarily concentrated east of the Andes near the MR. At longer lags ($-$12 hours to $-$18 hours), higher frequencies occurred progressively farther east, over the Colombian Orinoco basin, southwestern Venezuela, and northwestern Amazon basin. By $-$24 hours, the highest frequencies are located over the southwestern portion of the Guiana Highlands in Venezuela, a well$-$known hotspot of MCS activity in NOSA \citep{robledo2024climatological}. Extending the analysis to $-48$ h (not shown) produced a similar pattern, with the Guiana Highlands remaining the dominant region of convective activity. These findings suggest a persistent and recurrent source region, where convective precursors to MR MCSs can be traced back up to 24 hours in advance, providing a previously unquantified source of predictability for impacting MCSs.

\subsubsection{Synoptic regimes leading MCSs initiation and characteristics}

\begin{figure}[t]
 \noindent\centerline{\includegraphics[width=39pc,angle=0]{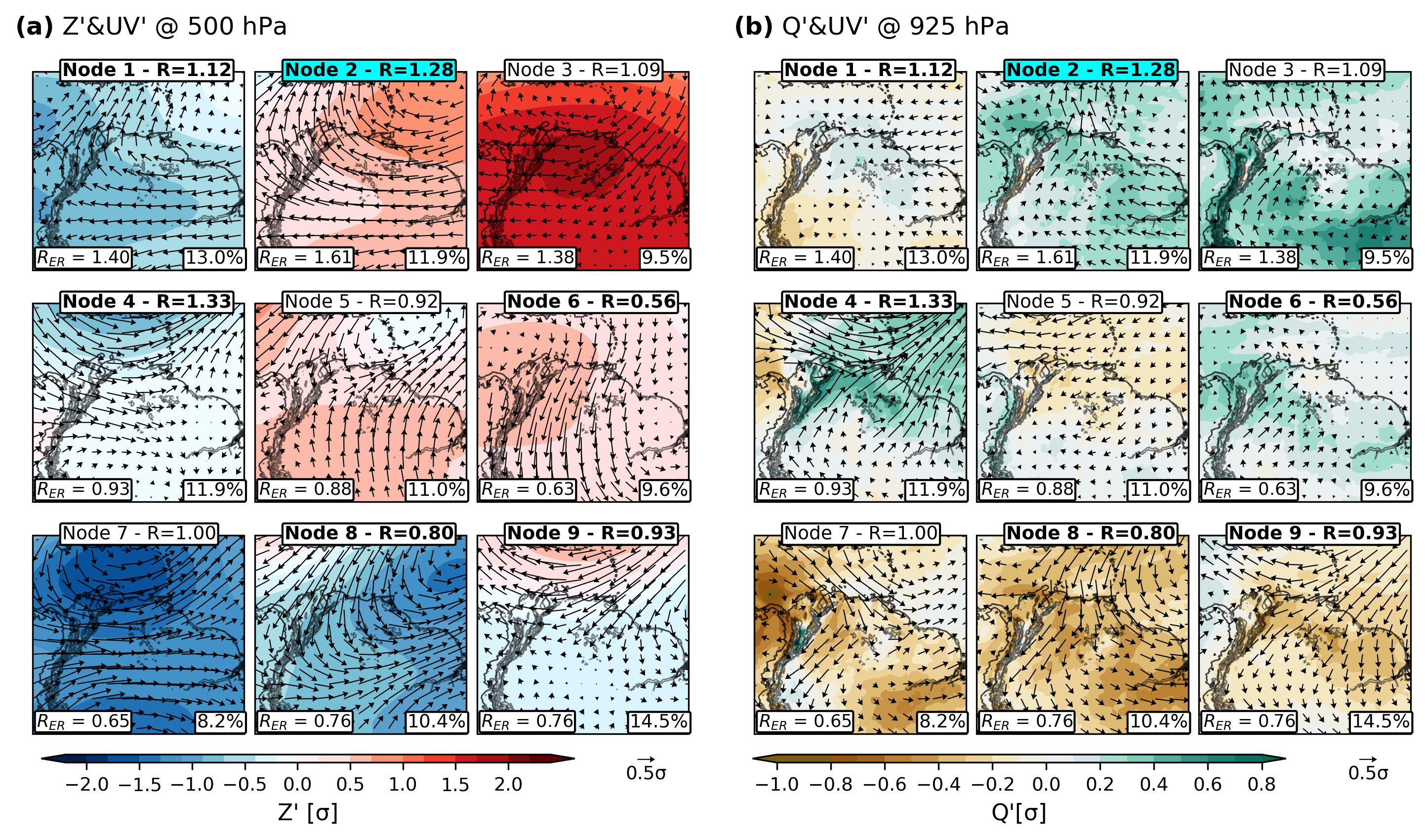}}\\
 \caption{Composite standardized anomalies of a) 500 hPa geopotential height and winds (vectors) and b) 925 hPa specific humidity and winds (vectors) for each large-scale environment identified by the SOM analysis during February--May. Enrichment Factor ($R$), where R$>$1 indicates favorable nodes for MCS initiation in the Orinoco--Amazon basin. Bold values indicate nodes with statistically significant $R$ at the 99\% confidence level based on a 10,000-iteration permutation test. The ER wave-conditioned enrichment factor ($R_{ER_i}$; lower left), indicates preferential MCS occurrence during active ER phases when $R_{ER_i}>1$. Percentages in the lower right indicate the climatological frequency of each SOM node. The cyan box identifies the node associated with the Mocoa event.}\label{f7}
\end{figure}

To answer whether the large-scale atmospheric environment associated with the Mocoa MCS was unique or anomalous relative to the climatology, we applied a Self-Organizing Map (SOM) analysis. \textbf{Figure \ref{f7}} presents the nine synoptic configurations identified during February--May (2001--2021) period. Each node represents a recurrent large-scale atmospheric configuration derived from the climatology, with the percentage in the lower-right corner indicating its frequency of occurrence. The SOM was trained using the same variables presented in \textbf{Figure \ref{f2}} to ensure consistency with the analysis of the Mocoa MCS. We quantified the degree of favorability for MCS initiation by computing the Enrichment Factor (R; see Section \ref{sec2} for details), where R$>$1 indicates atmospheric configurations associated with enhanced MCS occurrence, while R$<$1 indicates configurations associated with suppressed MCS activity. Statistically significant nodes are highlighted using the bold labels in \textbf{Figure \ref{f7}.}

Given the presence of an ER wave during the initiation of the Mocoa event, we also quantified the relative preference of MCS occurrence within each node during periods when ER waves were active. Similar to the Enrichment Factor, we defined an ER wave-conditioned enrichment factor $R_{ER_i}$ as:

\[
R_{ER_i} = \frac{P(node_i | (MCS, ER)}{P(node_i | (NoMCS, ER)}
\]

Values of $R_{ER_i} > 1$ indicate that the large-scale environment represented by the node is preferentially associated with MCS occurrence during active ER phases, whereas $R_{ER_i} < 1$ indicates suppressed MCS occurrence despite ER activity. Although the SOM analysis does not establish causality between large-scale regimes and MCS occurrence, below we provide associations of the structure and dynamic fields that support the interpretation of the nodes as analog regimes that could sustain/inhibit MCSs.

\begin{figure}[t]
\noindent\centerline{\includegraphics[width=39pc,angle=0]{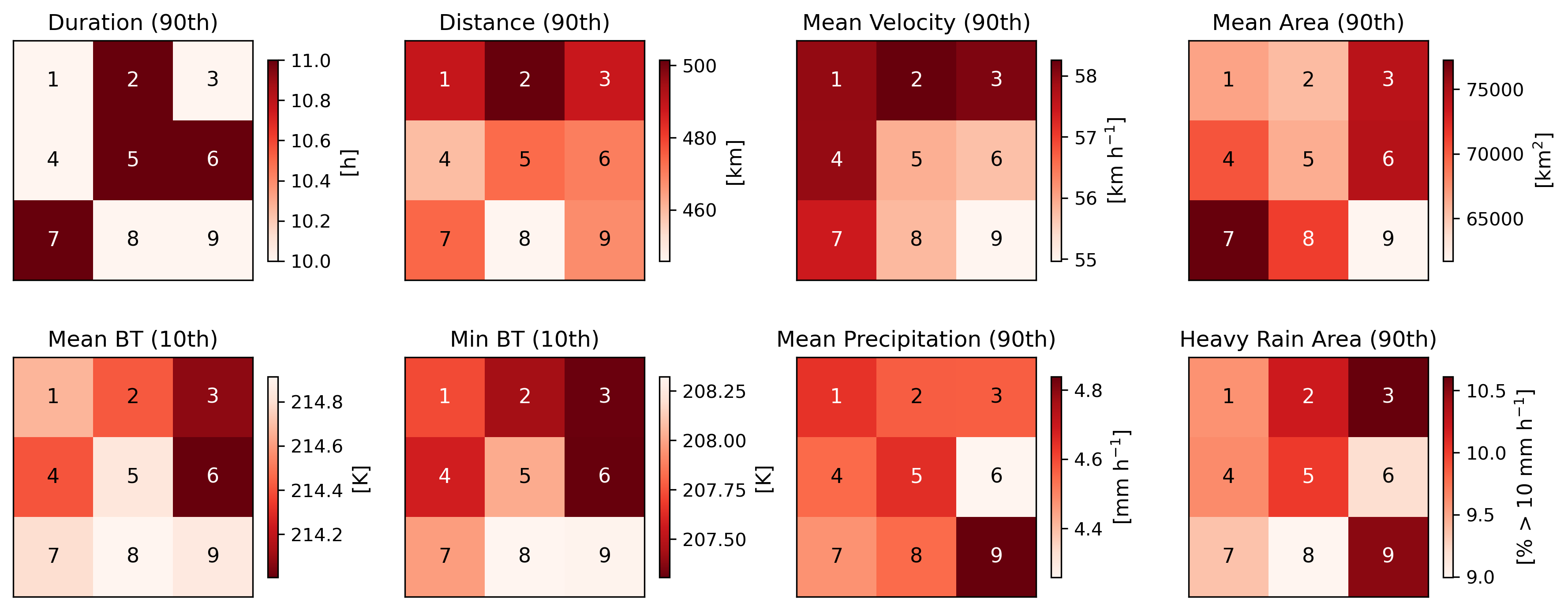}}\\
\caption{90$^{th}$ (and 10$^{th}$ for brightness temperature (BT)) percentile MCS characteristics across the nine SOM nodes. Panels (left to right, top to bottom) show duration, travel distance, mean propagation velocity, mean area, mean brightness temperature, minimum brightness temperature, mean precipitation rate, and heavy-rain area (percentage of the MCS area with precipitation rates $>$10 mm h$^{-1}$). Darker colors indicate larger values for the corresponding percentile.}\label{f8}
\end{figure}

When the Mocoa event environment was projected onto the SOM, it mapped to Node 2, identifying this node as the closest analog to the observed synoptic conditions (\textbf{Figure \ref{f2}}). At 500 hPa, Node 2 exhibits the previously identified ER-like pattern, characterized by the westward propagating positive geopotential height anomalies accompanied by anticyclonic wind anomalies. At 925 hPa, positive specific humidity anomalies coexist with westward wind anomalies over the Amazon basin and a weak OLLJ, represented by northward wind anomalies between the Andes and the Guiana Highlands. 
This configuration favors large-scale ascent by promoting divergence aloft over the Orinoco--Amazon basins as the flow deflects around the anomalous highs. The resulting ascent, combined with moisture accumulation likely associated with the weak OLLJ, enhances convective instability and provides favorable dynamical conditions for MCS initiation and maintenance. This interpretation is consistent with the relatively high enrichment factor (R=1.28). The association with the ER-like pattern is further supported by Node 2 having the highest ER wave-conditioned enrichment factor ($R_{ER_2}=1.61$). MCSs associated with Node 2 are also the longest-lived, fastest-moving, and farthest-traveling among all nodes (see \textbf{Figure \ref{f8}}), likely promoted by the westward movement of the ER wave-like pattern, the enhanced westward equatorial flow, and a wind profile suggestive of favorable vertical wind shear (see Figure\ref{fA7}) for sustained convective organization and propagation.

In addition to Node 2, five other nodes significantly favored or inhibited MCSs (bold labels in \textbf{Figure \ref{f7}a}). Nodes 1 and 4 were also identified as MCS--favorable regimes ($R=1.12$ and $R=1.33$, respectively), whereas Nodes 6, 8, and 9 were associated with MCS--inhibiting regimes.

Node 1 is characterized by weak negative geopotential height anomalies located over the northeastern and southeastern portions of the domain, coinciding with weak meridional geopotential height gradients in the raw fields (see Figure \ref{fA6}). The resulting circulation is anticyclonic over the northern region, with strong westward equatorial winds deflected around the Andes, resembling the mid-level flow pattern observed in Node 2.
At 925 hPa, weak negative specific humidity anomalies dominate the Amazon basin, indicating drier-than-normal conditions, whereas weak positive anomalies persist over the Orinoco basin (\textbf{Figure \ref{f7}b}). The positive moisture anomalies are primarily sustained by the westward flow entering from the Venezuelan coast. A neutral OLLJ is also evident, represented by weak anomalies in the wind magnitude along the eastern slopes of the Andes. 
As in Node 2, the anticyclonic circulation supports divergence and large-scale ascent, consistent with convective development and coinciding with hosting fast-moving and long-traveled MCSs (Figure \ref{f8}). However, its lower enrichment factor may reflect the reduced moisture availability over the Orinoco--Amazon basin, potentially limiting MCS longevity. Nevertheless, the relatively high ER wave-conditioned enrichment factor $R_{ER_1}=1.40$ suggests that ER waves remain an important dynamical contributor to MCS favorability within this regime.

Node 4 exhibits pronounced negative geopotential height anomalies at 500 hPa centered over the Caribbean Sea, while negative anomalies over the continental region remain comparatively weak. Wind anomalies are strongest over the northern and western portions of the domain. Over the Caribbean Sea, eastward wind anomalies reinforce the cyclonic circulation, resembling a trough-like pattern. In contrast, westward flow dominates over the continent (see Figure \ref{fA6}), but opposite direction of the wind anomalies indicate a weakening of the mid-level easterlies, likely associated with the intensified eastward flow linked to the Caribbean trough-like circulation. 
At 925 hPa, positive specific humidity anomalies are concentrated over the Orinoco basin and the Caribbean region. Eastward wind anomalies over the Caribbean Sea indicate a weakened Caribbean Low-Level Jet (CLLJ; \citep{maldonado2016CLLJ}), and northeastward wind anomalies along the eastern flank of the Andes indicate a weak OLLJ. 

Similar to a mid-latitude trough, the cyclonic circulation likely enhances convection by promoting upper-level divergence and positive vorticity advection along its eastern-equatorward flank, supportive of large-scale ascent. Negative anomalies of geopotential height are usually tied to colder mid-level temperatures, which, over a warm and moisture-rich boundary layer, steepen the environmental lapse rate, reducing convective inhibition (CIN), and increasing convective available potential energy (CAPE), making it much easier for air parcels to achieve deep, buoyant ascent \citep{stull2012introduction}. This interpretation is consistent with the cold cloud tops depicted in \textbf{Figure \ref{f8}}. Together with the positive low-level humidity anomalies, these conditions likely create a favorable environment for MCSs and may explain why Node 4 exhibits the highest enrichment factor among all nodes (R=1.33), despite its weak association with ER waves ($R_{ER_4}=0.94$). 
The contrast between the mid- and low-level winds also suggests enhanced directional vertical wind shear over the northern portion of the domain. Over the Orinoco--Amazon basin, this contrast is not evident, and vertical wind shear is lower compared to Node 1 and Node 2 (see Figure \ref{fA7}). As stated by \citet{rotunno1988theory}, if vertical wind shear is either too low or too high, the MCS structurally uncouples and dies. Consistent with this interpretation, Node 4 is associated with the shortest-lived MCSs among all nodes (see \textbf{Figure \ref{f8}}). 

Node 6 represents an MCS--inhibiting regime (R=0.56). Positive geopotential height anomalies dominate the mid-troposphere, with meridional gradients north of Colombia, consistent with a more stable environment and enhanced subsidence (\textbf{Figure \ref{f7}a}). At 500 hPa, pronounced southerly wind anomalies extend from the Caribbean Sea into the continental region, consistent with the southeasterly flow evident in the raw wind field (Figure \ref{fA6}). At 925 hPa, positive specific humidity anomalies are concentrated over the Colombian Caribbean region and east of the Colombian Andes. The low-level wind anomalies indicate a weakened OLLJ, while the vertical wind shear remains enhanced over the OLLJ region (Figure \ref{fA7}).This combination may help explain why the relatively few MCSs that develop within this otherwise unfavorable regime tend to be long lived but exhibiting the lowest mean precipitation rates among all nodes (\textbf{Figure \ref{f8}}). Node 6 also exhibits the weakest association with ER-wave activity among the nine nodes ($R_{ER_4}=0.94$) which may further limit the large-scale support for MCS occurrence and contribute to its low enrichment factor. 

Node 8 at 500 hPa is characterized by two negative geopotential height anomalies accompanied by strong cyclonic wind anomalies near the equatorial region (\textbf{Figure \ref{f7}a}). The geopotential height field exhibits a strong zonal gradient over the Caribbean Sea, consistent with the positive geopotential height anomalies over the northwestern portion of the domain. This configuration is conducive to northerly flow into the continent that converges with the easterlies over the Orinoco basin (Figure \ref{fA6}), potentially enhancing mid-level convergence and low-level divergence, reducing convective activity.
At 925 hPa, the node exhibits negative specific humidity anomalies together with wind anomalies indicating a strengthened OLLJ. The reduced low-level moisture availability and the strengthened OLLJ may support rapid moisture transport through the region rather than moisture accumulation and local convective initiation, contributing to the MCS-inhibiting character of this regime. MCSs associated with this node tend to be short-lived, propagate over shorter distances, and have lower heavy rain areas compared to other nodes. 

Finally, Node 9 exhibits a circulation pattern approximately opposite to Node 4, with positive geopotential height anomalies centered over the Caribbean Sea and enhanced anticyclonic flow, supported by both the wind anomalies (\textbf{Figure \ref{f7}a}) and the raw wind fields in Figure \ref{fA6}. At 925 hPa (\textbf{Figure \ref{f7}b}), negative specific humidity anomalies coexist with easterly and southeasterly wind anomalies over the Caribbean Sea and Orinoco--Amazon basin, respectively. These anomalies are consistent with stronger-than-normal CLLJ and OLLJ. Similar to node 8, this configuration may support enhanced low-level ventilation, limiting moisture accumulation and sustained moisture convergence, contributing to its MCS-inhibiting character (R=0.93). MCSs developing within this regime are generally short-lived and small but exhibit relatively high precipitation rates (Figure \ref{f8}).

\begin{figure}[t]
\noindent\centerline{\includegraphics[width=19pc,angle=0]{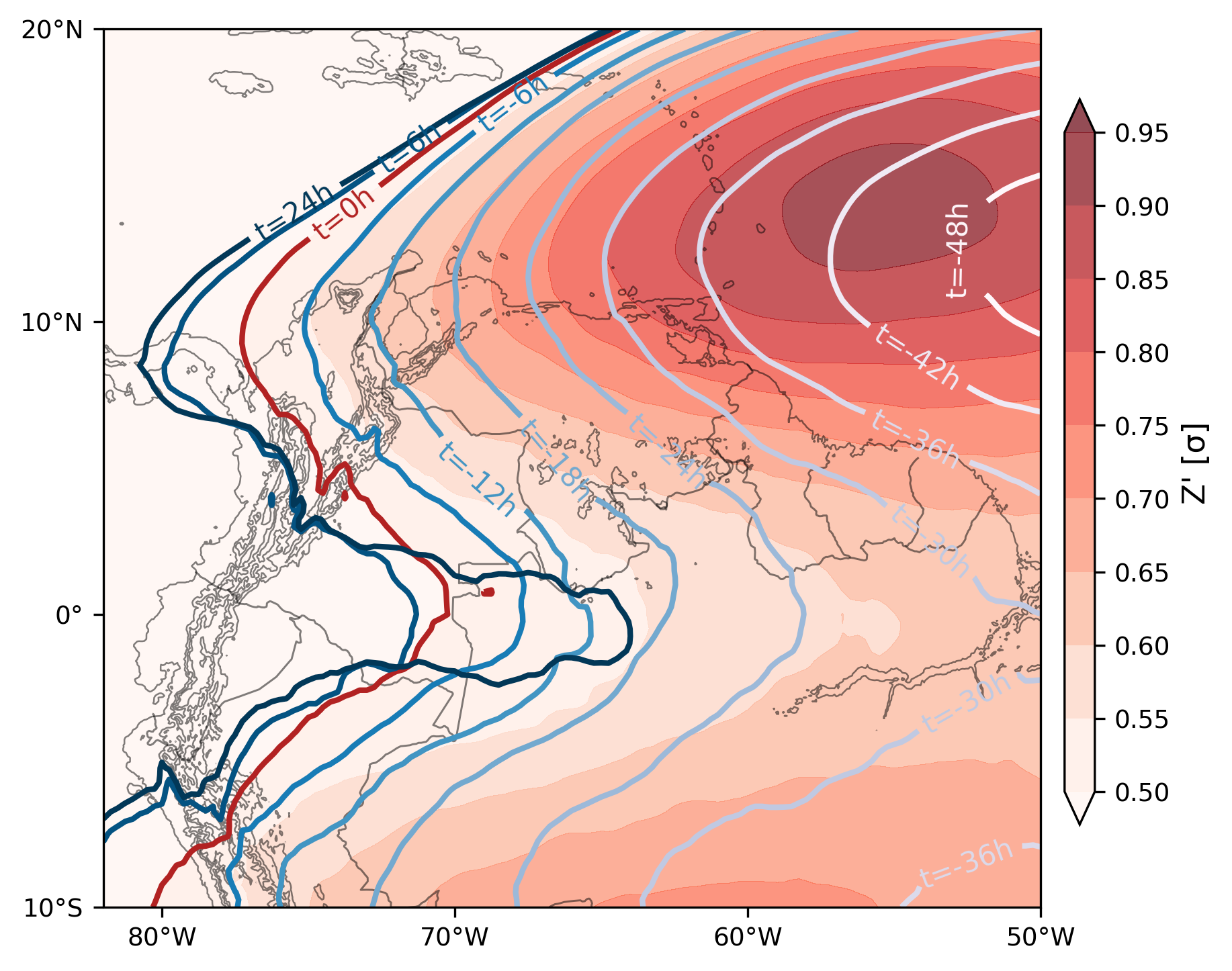}}\\
\caption{Lag-composite of standardized 500-hPa geopotential height anomalies for all MCS events classified within SOM node \#2. Filled contours show the composite anomaly field at MCS initiation. Overlaying isolines ($Z^\prime = 0.5$) depict the westward propagation of the composite anomaly from 48 h before MCS initiation (t= $-48$h) to 24 h afterward (t=$24$ h).}\label{f9}
\end{figure}

Given our focus on Node 2, we further evaluated whether the westward-propagating signal identified in the Mocoa MCS composites (Section 3a) was also evident in the broader population of MCSs associated with this node. \textbf{Figure \ref{f9}} presents a lag-composite of all MCS environments classified within Node 2, spanning from 48 h before MCS initiation through the early stages of development. The results reveal a coherent westward propagation of the ER-like pattern, consistent with the characteristic phase propagation of ER waves. As MCS initiation approaches, the meridional geopotential height gradient becomes more pronounced as the positive anomalies north and south of the equator strengthen. At $t = -48h$, the anomaly pattern is located over the tropical Atlantic northeast of South America, and progressively propagates westward toward the continent as MCS initiation approaches, with the isolines transitioning from white to darker blue. As the pattern reaches the Andes foothills near $t = +6h$, its westward propagation slows considerably. Previous studies have shown that topography and continental landmasses can substantially modify Rossby wave phase speeds and even induce blocking \citep{berbery1989observational, yang2015interaction, yang2021generalized}. This mechanism may explain the reduced propagation speed of the ER-like pattern observed between t = +6 h and t = +24 h. However, evaluating the extent to which Andean topography modulates the propagation of this ER-like pattern requires further investigation.

\subsubsection{The role of the Orinoco-Low Level Jet in MCS occurrence}

\begin{figure}[t]
\noindent\centerline{\includegraphics[width=39pc,angle=0]{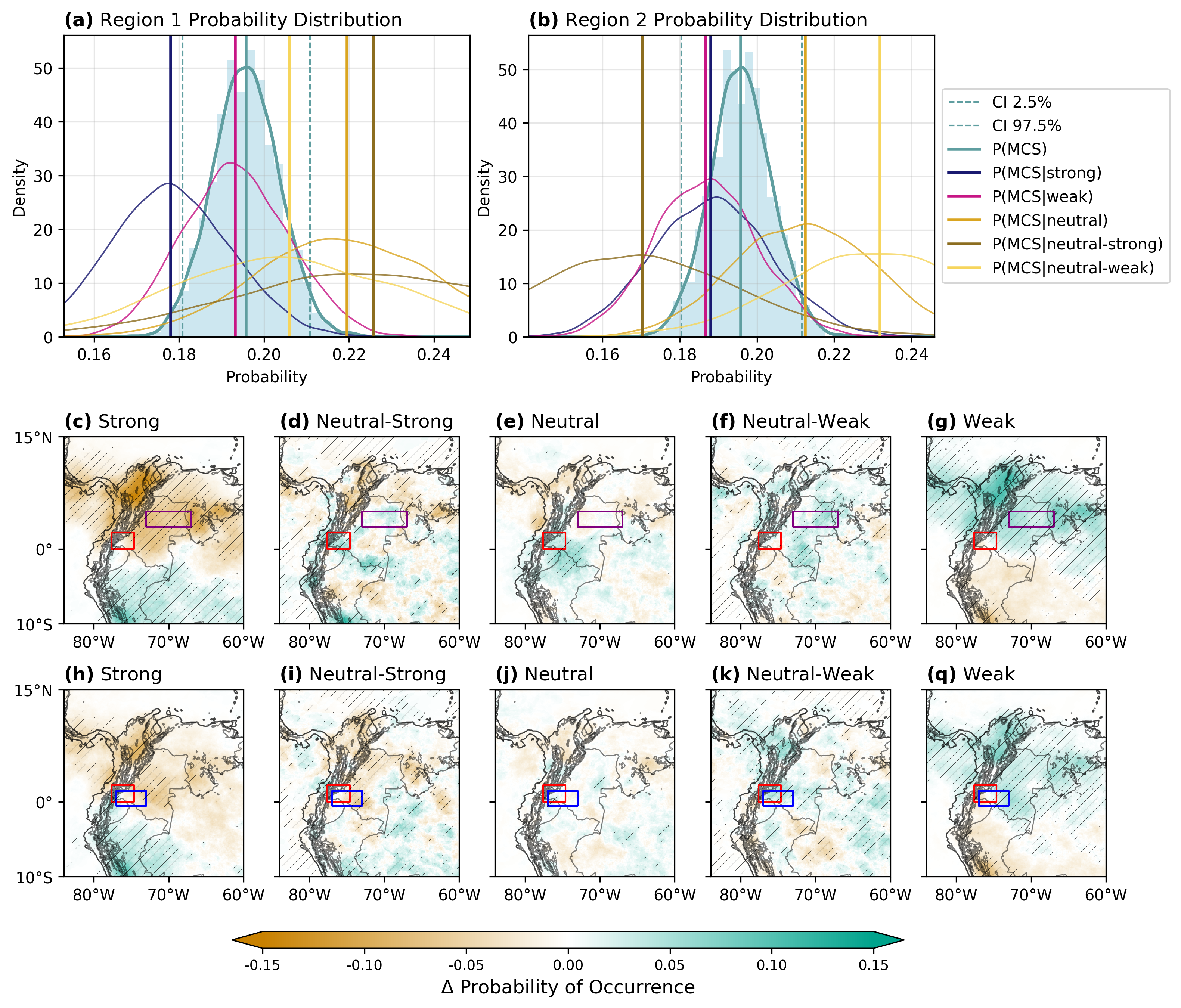}}\\
\caption{Relationship between OLLJ index and MCS occurrence. (a,b) Probability of MCS occurrence in the Mocoa region (red box in panels c--q) as a function of the OLLJ state in Region 1 (purple box; panel a) and Region 2 (blue box; panel b). Panels (c--g) and (h--q) show the change in MCS occurrence probability associated with different OLLJ states in Region 1 and Region 2, respectively. OLLJ states are defined as Strong ($\mathrm{OLLJ}\geq1$), Neutral--Strong ($0.5\leq\mathrm{OLLJ}<1$), Neutral ($-0.5<\mathrm{OLLJ}<0.5$), Neutral--Weak ($-1<\mathrm{OLLJ}\leq-0.5$), and Weak ($\mathrm{OLLJ}\leq-1$). Hatched areas are statistically significant at 99\% confidence level based on 1,000 bootstrap random iterations.}\label{f10}
\end{figure}

As discussed in \textbf{Section \ref{3_a_3}a}, the Mocoa MCS intensified under an active but climatologically neutral OLLJ. Consistent with this, the SOM analysis revealed that MCS--favorable large-scale environments were generally associated with neutral to weak OLLJ conditions, whereas MCS--inhibiting nodes were associated by stronger jets. Here, we further examine the relationship between OLLJ intensity and MCS occurrence by subdividing the OLLJ index into five intensity classes to better understand its influence over the MR. (\textbf{Figure \ref{f10}}).

When the OLLJ index was evaluated over Region 1 (\textbf{Figure \ref{f10}a}), corresponding to the core of the Orinoco basin, strong jet conditions ($\mathrm{OLLJ}\geq1\sigma$) were associated with a statistically significant reduction in MCS occurrence over the MR relative to climatology. Weak jet conditions also showed below-climatological probabilities, although the reduction was smaller and not statistically significant. In contrast, Neutral--Strong ($0.5\leq\mathrm{OLLJ}<1$) and Neutral ($-0.5<\mathrm{OLLJ}<0.5$) conditions showed the highest MCS occurrence probabilities over the MR, both significantly exceeding climatology.

The spatial distributions (\textbf{Figures \ref{f10}c--g}) show how OLLJ intensity modulated the preferred regions of MCS occurrence. During strong jet conditions, MCS occurrence decreased by up to 15\% across Colombia, Venezuela, and northern Brazil, while increasing over Peru and southwestern Amazonia. Conversely, weak jet conditions shifted enhanced MCS occurrence across northern NOSA. The highest probabilities near southwestern Colombia (close to MR) occurred under Neutral and Neutral--Strong jet conditions. These patterns suggest that a strong jet may advect moisture rapidly downstream, displacing convection away from the MR, whereas moderate jet conditions may favor moisture accumulation and convective organization near the southeastern Andes foothills.

Evaluating the OLLJ over Region 2 (\textbf{Figure \ref{f10}b}), located near the jet exit region and close to MR, revealed a somewhat different relationship but same physical mechanism. Strong, Neutral--Strong, and Weak jet conditions were associated with reduced MCS occurrence over the MR, with statistically significant suppression in the Neutral--Strong category. In contrast, Neutral and Neutral--Weak conditions significantly increased MCS occurrence over the MR. The spatial distributions (\textbf{Figures \ref{f10} h-q}) indicate that Neutral--Weak conditions preferentially enhanced MCS occurrence near the Orinoco--Amazon transition and the jet exit region, whereas Weak conditions favored convection more broadly across Colombia and Venezuela.

\section{Summary and conclusions}

We investigated the atmospheric conditions that led to the catastrophic March 2017 Mocoa MCS-induced flash flood and examined whether its environment corresponded to recurrent large-scale configurations associated with MCS occurrence in the Orinoco--Amazon basin. Using a satellite-based MCS climatology, reanalysis data, and SOM analysis, we characterized the large-scale processes associated with the event and evaluated their relationship with climatological patterns of MCS development.

When compared against the regional MCSs climatology affecting the MR, the March 2017 MCS exhibited both typical and exceptional characteristics. The MCS developed during the climatologically favored season and followed the dominant westward propagation pathway across the Orinoco--Amazon basin. However, it initiated farther east and earlier in the day than typically observed, while ranking among the largest, longest-lived, and fastest-propagating MCSs in the climatological record. Composite analyses further revealed an upstream source region over the Guiana Highlands with enhanced probabilities of MCS initiation up to 24 hours before MCS arrival in the MR. The initiation location of the Mocoa MCS coincided with one of these preferred upstream regions, highlighting the Guiana Highlands as a recurrent source region for MCSs affecting southern Colombia. These results suggest that convective precursors to MCSs impacting the southern Colombian Andes may be identified well before the systems reach the eastern Andes foothills. The role of this upstream source region in modulating hazardous convection over the eastern Andes deserves further investigation. 

The SOM analysis identified nine recurrent large-scale atmospheric configurations (nodes) associated with distinct frequencies of MCS occurrence across the Orinoco--Amazon basin. Among them, the environment associated with the Mocoa event, characterized by a westward-propagating ER-like pattern, enhanced low-level moisture, a neutral OLLJ, and favorable upper-level circulation, exhibiting an enrichment factor of $R=1.28$, the highest ER-wave-conditioned enrichment factor ($R_{ER}=1.61$), and a climatological frequency of approximately 11\%. MCSs associated with this node were also longer lived, propagated faster, and traveled farther han those associated with the remaining SOM nodes, characteristics also exhibited by the Mocoa MCS. These findings suggest that the large-scale atmospheric regimes identified here could modulate both the likelihood of MCS initiation and the characteristics of the MCSs that develop within them, providing an additional pathway through which synoptic-scale processes can contribute to MCS predictability.

The OLLJ emerged as another important component of these favorable environments. Consistent with \citet{martinez2024mesoscale}, we identified an active OLLJ during the intensification of the Mocoa event, although our results indicated neutral-to-moderately enhanced rather than exceptionally strong jet conditions. More broadly, the climatological analysis revealed that weaker OLLJ conditions favored MCS development across northern NOSA, whereas neutral-to-moderately enhanced conditions maximized MCS occurrence near the jet exit region and the MR. In contrast, stronger OLLJ conditions shifted convective activity southward across NOSA which is consistent with results reported by \citet{mu2025impacts}. These results suggest that balanced moisture transport, rather than maximum low-level flow intensity, provides a more favorable environment for convective organization near the eastern Andes foothills. They further indicate that accurately representing low-level jet variability, including its seasonal and diurnal evolution, may provide additional predictive information for hazardous convection in the region.

Overall, MCS--favorable environments were consistently characterized by enhanced low-level moisture, neutral-to-moderately OLLJ conditions, favorable upper-level circulation, and increased ER wave activity. In contrast, MCS--inhibiting environments generally exhibited reduced moisture availability, stronger low-level jets, and weaker ER wave influence. While our results do not establish a causal relationship between ER waves  and individual MCS events, they provide evidence that ER waves, acting together with favorable thermodynamic and dynamical conditions, are associated with an increased likelihood of organized convection. This interpretation is consistent with previous studies documenting enhanced MCS occurrence and extreme rainfall during ER-wave phases \citep{cheng2023mesoscale,king2017mechanisms,nakamura2022aconvective,nakamura2022bconvective}, while extending those findings by placing ER-wave activity within a broader synoptic environment conducive to MCS development. 

Taken together, our results support the hypothesis that recurrent large-scale atmospheric configurations create preferred environments for MCS development in the Orinoco--Amazon basin. The atmospheric conditions associated with the catastrophic Mocoa MCS event were not exceptional from a climatological perspective, but rather represented a recurrent, yet previously undocumented, preferential state for convective organization. Because these recurrent large-scale patterns can be identified and monitored several hours to days in advance, they may provide forecasts of opportunity by tracking the atmosphere toward states of enhanced convective probability. Identifying these favorable synoptic regimes and improving their representation in operational forecasting systems may ultimately help extend warning lead times for MCS-driven hazards and flash floods in northern South America and other tropical regions.

Additionally, some limitations present in this work could merit future work, including extending the analysis to other seasons apart from the February–May period, and further assessing the sensitivity of our results to uncertainties associated to relying on reanalysis products and the SOM classification. The recurrent synoptic environments identified here provide a framework that can be tested, refined, and extended as additional observations and forecasting systems become available. Future work should include extending the analysis to other seasons beyond February-May, quantify the predictive skill of the identified synoptic regimes and evaluate the ability of operational forecasting systems to represent the identified precursors.

\clearpage
\acknowledgments

Vanessa Robledo was supported by the NASA Earth and Space Science and Technology (FINESST) grant 80NSSC25K0649.  Mejia was partially funded by the Division of Atmospheric Sciences, Desert Research Institute.

%
%
\datastatement

The MCSs tracking database is available at \citet{robledo_delgado_2024_10443552}. The ERA5 dataset can be obtained from \url{https://cds.climate.copernicus.eu/#!/search?text=ERA5&type=dataset}. SOM is run by the Python miniSOM library: \url{https://github.com/JustGlowing/minisom}.

%



\appendix[A]\label{AA}


\appendixtitle{SOM sensitivity analysis}

\begin{figure}[h]
\noindent\centerline{\includegraphics[width=27pc,angle=0]{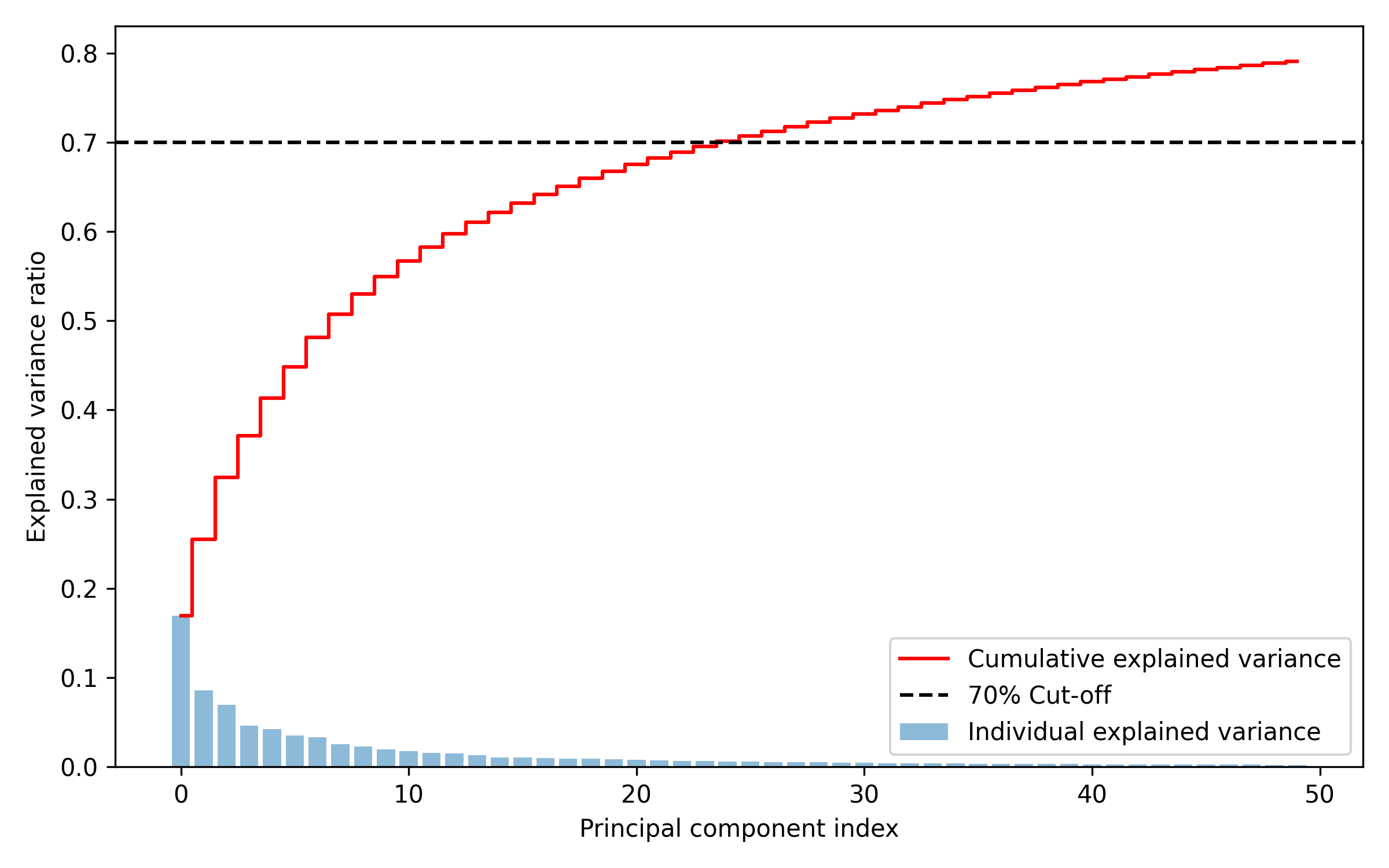}}\\
\caption{Explained variance by components.}\label{fA1}
\end{figure}

Figure \ref{fA1} shows Principal Component Analysis (PCA) scree plot displaying the explained variance ratio across 50 principal components.

Following previous SOM applications in synoptic meteorology  \citep{cassano2016synoptic, wang2022linking, ford2015synoptic, mechem2018joint}, the SOM configuration was selected through a sensitivity analysis rather than prescribed a priori. Because the number of SOM nodes determines the level of detail represented by the classification, too few nodes may merge distinct circulation regimes, whereas too many may produce redundant patterns with fewer samples per node, reducing robustness and interpretability.

\begin{figure}[h]
\noindent\centerline{\includegraphics[width=33pc,angle=0]{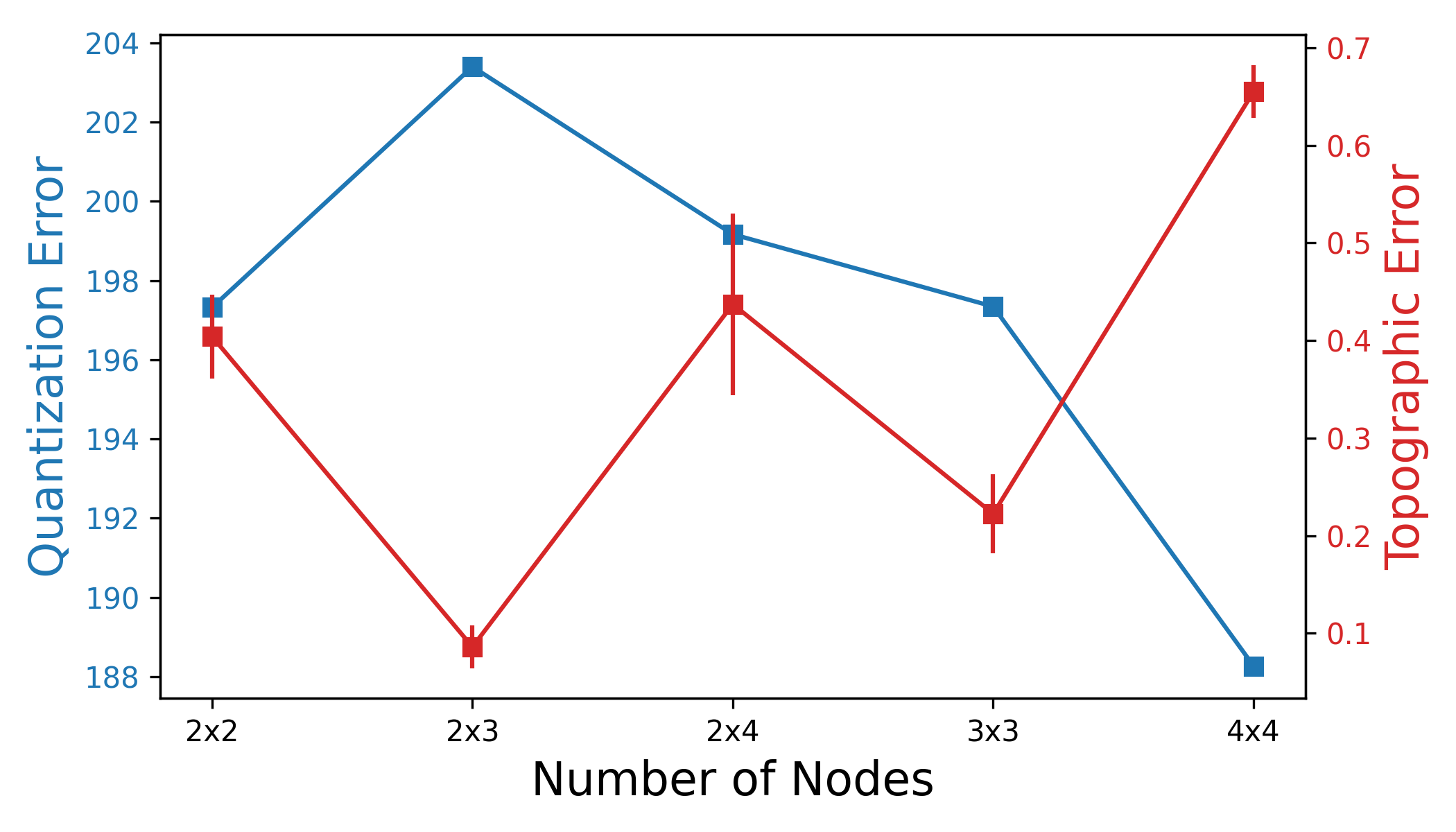}}\\
\caption{Sensitivity of Self-Organizing Map (SOM) performance to grid size. The mean quantization error (QE; blue, left axis) and topographic error (TE; red, right axis) are shown for different SOM configurations, with error bars representing variability across random initializations.}\label{fA2}
\end{figure}

We evaluated five map configurations (2×2, 2×3, 2×4, 3×3, and 4×4), five random initialization seeds (2, 4, 10, 20, and 40), seven neighborhood radii (sigma= 0.3, 0.5, 0.8, 1.0, 1.2, 1.5, and 2.0), and four learning rates (0.05, 0.10, 0.30, and 0.50), using 60,000 training iterations for each experiment. For each map size, the hyperparameter combination yielding the lowest QE and TE was identified. The selected configuration was then retrained using the different random initialization seeds, and the mean and standard deviation of the quantization error (QE) and topographic error (TE) were computed across these independent realizations. QE measures the distance between each data sample and its winning neuron (BMU: Best Matching Unit), with smaller values indicating better representation of the input space \citep{kohonen2013essentials}. TE is the average geometric distance between the winning and the second-best matching nodes, it evaluates how well neighborhood relationships among similar patterns are preserved on the SOM grid \citep{doan2021s, wang2022linking}. The hyperparameter combination with the lowest QE and TE was used for the comparison across grid sizes (Figure \ref{fA2}).

Figure \ref{fA2} shows that increasing the number of nodes generally reduced QE by improving representation of the input space but increased TE, reflecting reduced topology preservation. This trade-off has been noted in previous atmospheric SOM studies \citep[e.g.,][]{wang2022linking}. The 3×3 configuration provided the best balance between representation accuracy, topology preservation, stability across random initializations, and the identification of distinct yet interpretable circulation regimes, and was therefore selected for the analyses. 

\appendix[B]\label{AB}
\appendixtitle{Mocoa MCS characteristics and environment}

\begin{figure}[ht!]
\noindent\centerline{\includegraphics[width=19pc,angle=0]{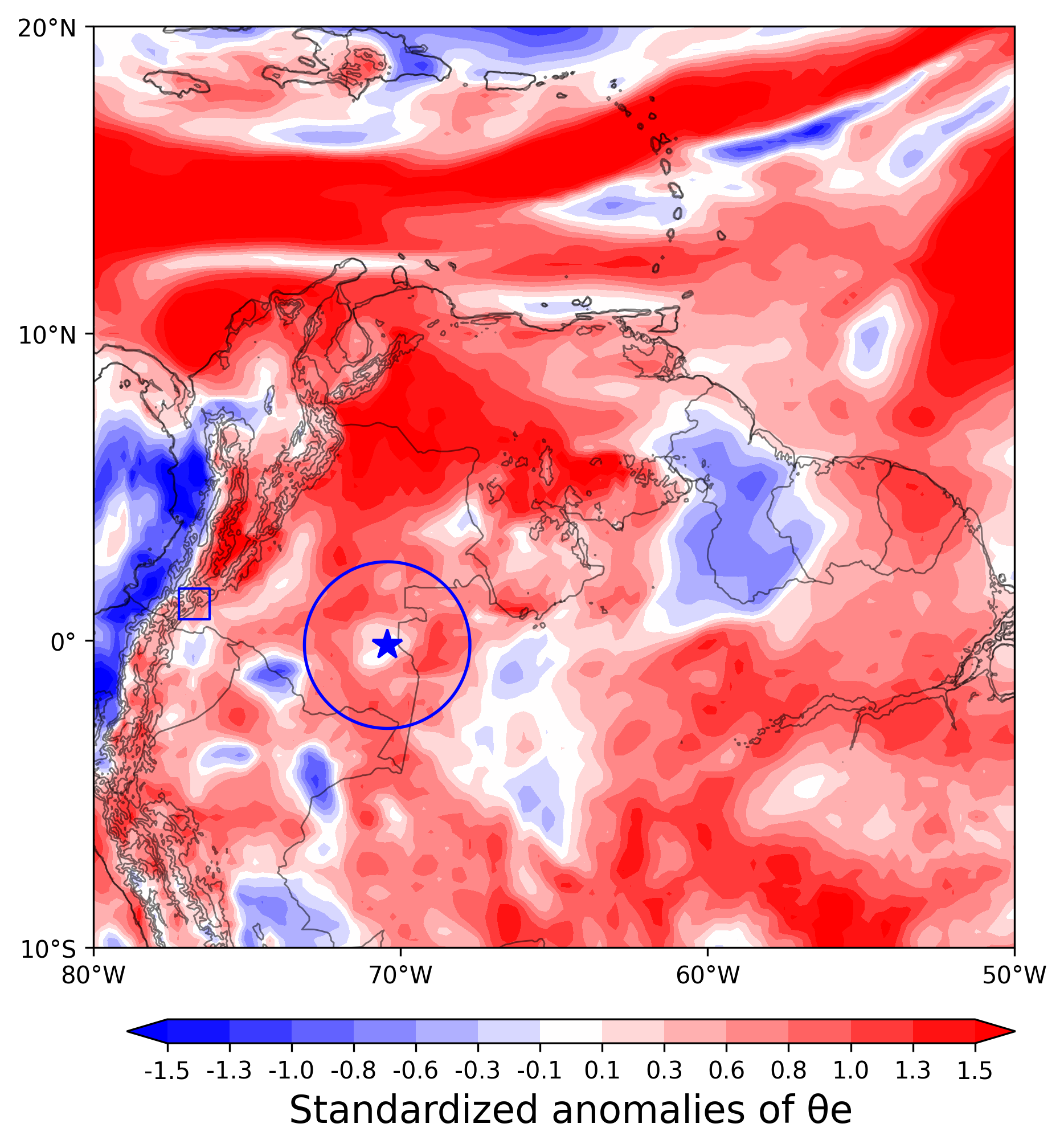}}\\
\caption{24-hour ($30^{th}$ March 1400 LT - 31$^{st}$ March 1400 LT) composites of standardized anomalies of equivalent potential temperature at 925 hPa. Blue star indicate Mocoa MCS initiation centroid.}\label{fA3}
\end{figure}

\begin{figure}[hb!]
\noindent\centerline{\includegraphics[width=19pc,angle=0]{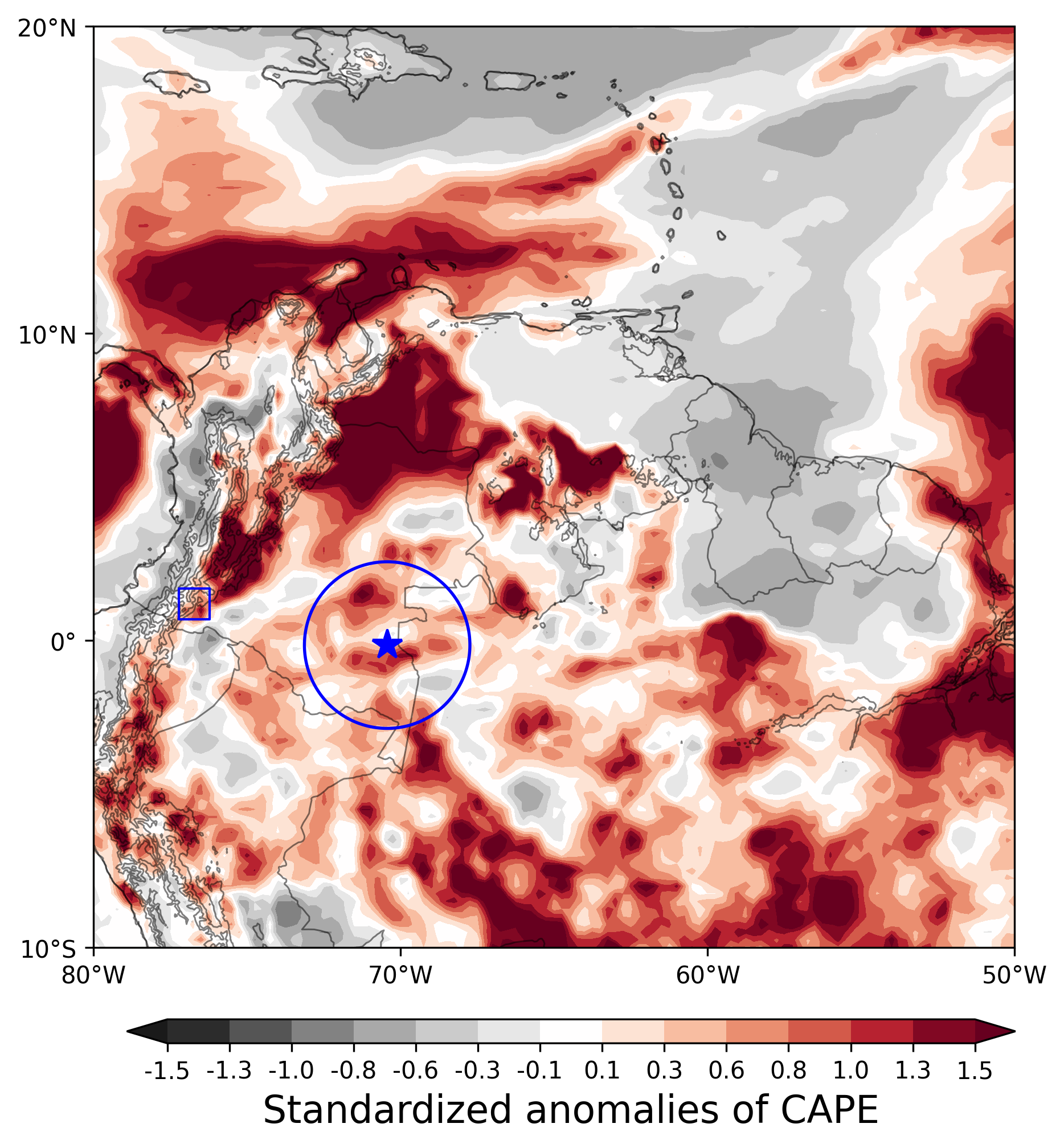}}\\
\caption{24-hour ($30^{th}$ March 1400 LT -- 31$^{st}$ March 1400 LT) composites of standardized anomalies of CAPE.  Blue star indicates Mocoa MCS initiation centroid.}\label{fA4}
\end{figure}

\begin{figure}[ht!]
\noindent\centerline{\includegraphics[width=27pc,angle=0]{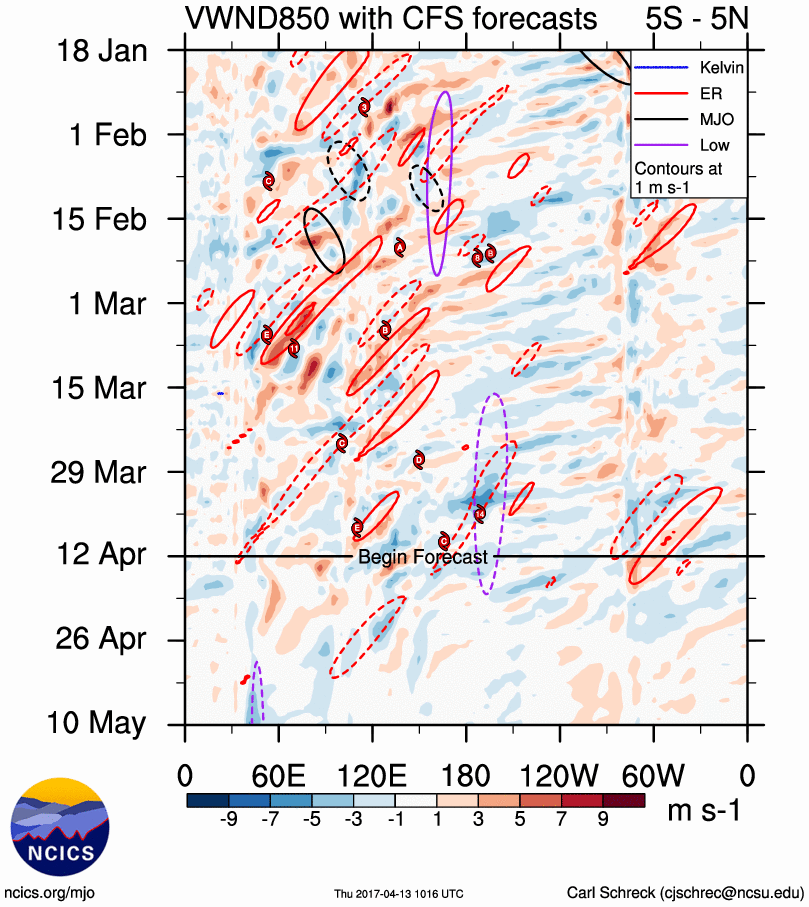}}\\
\caption{Longitude-time Hovmöller of meridional wind (vwnd850). Shading shows anomalies from climatology in the past. In the future, shading is either the sum of all modes or the CFS forecast and its departure from its own climatology. Contours identify the MJO, equatorial Rossby waves, Kelvin waves, and the low-frequency background ($>$ 120 days). Hurricane symbols denote tropical cyclogenesis identified by NHC and JTWC, with the direction indicating Northern or Southern Hemisphere. Image taken from https://ncics.org/pub/mjo/}\label{fA5}
\end{figure}

\begin{figure}[ht!]
\noindent\centerline{\includegraphics[width=39pc,angle=0]{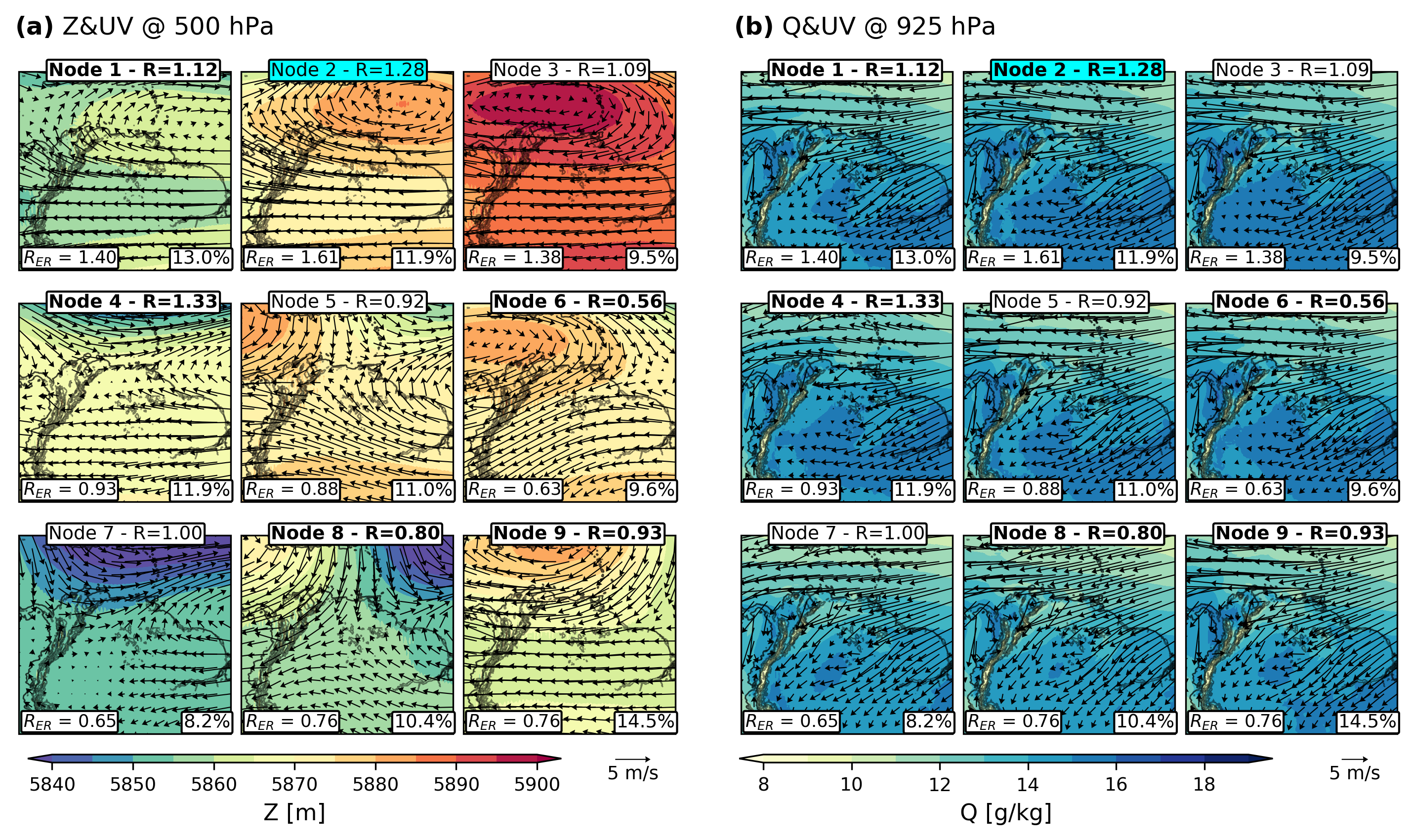}}\\
\caption{(a) Composites of 500-hPa winds (vectors; m s-1) and geopotential height (shading; gpm), (b) 925-hPa winds (vectors; m s-1) and specific humidity (shading; g kg-1) during February-May in nine nodes based on SOM analysis.}\label{fA6}\end{figure}

Table \ref{t1} summarizes the characteristics of the Mocoa MCS derived from the ATRACKCS algorithm. Figures \ref{fA3} and \ref{fA4} show composites of standardized $\theta_e$ and CAPE anomalies during the 24 h preceding MCS initiation. Figure \ref{fA5} presents the Equatorial Rossby (ER), Kelvin, Madden--Julian Oscillation (MJO), and low-frequency background signals from the North Carolina Institute for Climate Studies (NCICS). Figure \ref{fA6} shows the composite raw fields for the nine SOM nodes.

\begin{table}[h]
\caption{Characteristics of MOCOA MCS event according to ATRACKCS track.}\label{t1}
\begin{center}
\begin{tabular}{ccccrrcrc}
\topline
$Characteristics$ & $Value$ \\
\midline
 Duration (h) & 13 \\
 Mean velocity (km$h^{-1}$) & 65 \\
 Total distance traveled (km) & 3,181 \\
 Mean area (km$^{2}$) & 291,168 \\
 Mean precipitation rates (mm$h^{-1}$) & 5.14 \\
 Direction & west \\
\botline
\end{tabular}
\end{center}
\end{table}

\clearpage

\begin{figure}[ht!]
\noindent\centerline{\includegraphics[width=27pc,angle=0]{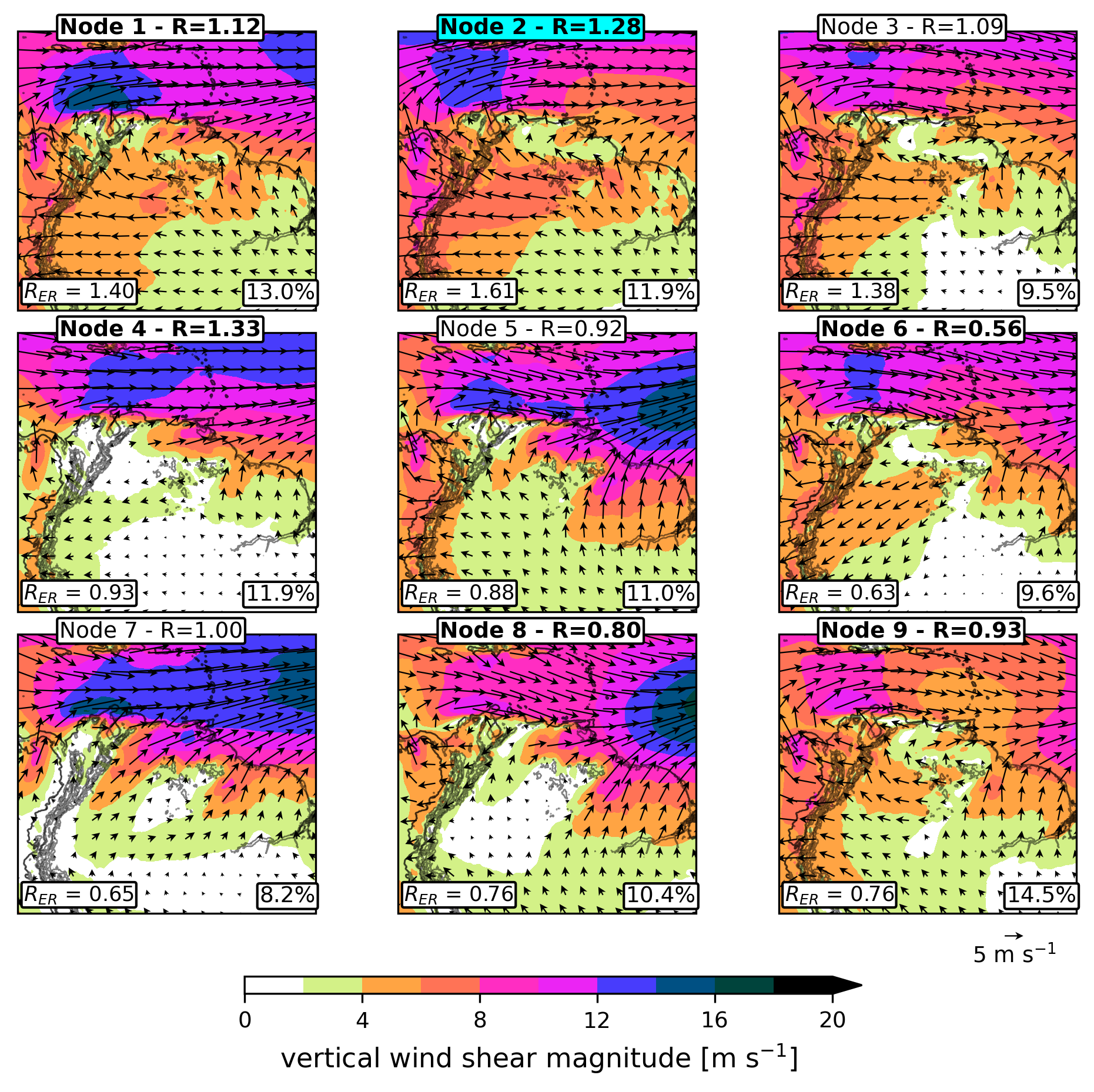}}\\
\caption{Composites of 500-925 hPa vertical wind shear direction  (vectors) and magnitude (shading; m s-1) during February-May in nine nodes based on SOM analysis.}\label{fA7}\end{figure}

\clearpage

\newpage

\bibliographystyle{ametsocV6}
\bibliography{references}

\end{document}